\documentclass{article}
\usepackage[preprint]{neurips_2024}
\usepackage[utf8]{inputenc}
\usepackage[T1]{fontenc}
\usepackage{amsmath,amssymb,amsfonts}
\usepackage{algorithm}
\usepackage{algpseudocode}
\usepackage{algorithmicx}
\usepackage{float}
\usepackage{booktabs}
\usepackage{array}
\usepackage{multirow}
\usepackage{enumitem}
\usepackage{graphicx}
\usepackage{caption}
\usepackage{subcaption}
\usepackage{xcolor}
\usepackage{url}
\usepackage{hyperref}
\usepackage[numbers,sort&compress]{natbib}

\newtheorem{theorem}{Theorem}

\title{Diffusion-guided optimization for full waveform inversion}

\author{%
Yiran Shen, Yangkang Chen, and Bj{\"o}rn Engquist
}

\begin{document}
\maketitle

\begin{abstract}
We present a diffusion-guided full waveform inversion (FWI) study in which pretrained diffusion generative models are used as learned regularizers inside a PDE-constrained seismic inversion loop. We compare three training-free guidance strategies: Manifold-Preserving Guided Diffusion (MPGD), SDEdit-based initialization, and Split Gibbs Diffusion Sampling (SGDS), which alternates between FWI likelihood updates and diffusion-prior denoising. The proposed workflow keeps wave-equation modeling in the inversion loop and uses a geological prior to stabilize model components that are weakly constrained by the seismic data. Controlled GeoFWI experiments, benchmark-scale Marmousi and Overthrust tests, a difficult Sigsbee2A salt test, and noise-degradation studies show that SGDS improves reconstruction quality relative to conventional $L_2$ and total-variation regularized FWI in clean and moderately noisy synthetic settings. Overall, these experiments demonstrate that diffusion-guided optimization can serve as a practical learned regularization strategy for synthetic FWI benchmarks while preserving the wave-equation modeling loop.
\end{abstract}

\section{Introduction}
Full waveform inversion (FWI) reconstructs subsurface physical properties by minimizing the mismatch between observed and simulated seismic wavefields. Because it exploits the full waveform information, FWI can achieve high spatial resolution compared with conventional imaging methods \cite[]{tarantola1984inversion, virieux2010overview, li2024robust}. By recovering high-resolution
velocity models, FWI provides critical inputs for hydrocarbon exploration, geothermal energy, and carbon storage monitoring. Despite its success, FWI remains fundamentally challenged by nonlinearity, ill-posedness, and sensitivity to noise and initial models. Recent efforts have explored learning-based strategies to improve inversion robustness, such as deep-learning matching filters that mitigate cycle skipping and improve convergence stability \cite[]{li2026dlm}. In particular, the lack of informative priors often leads to cycle-skipping and spurious artifacts that limit its applicability in complex geological settings.

Classical regularization strategies such as Tikhonov (or Sobolev) regularization trace back to the foundational work by Tikhonov (1963) \cite[]{engl2015regularization}, which introduced the idea of penalizing solution norms to stabilize ill-posed problems. For edge preservation reconstruction, Rudin-Osher-Fatemi (1992) popularized the total variation regularization \cite[]{rudin1992nonlinear}, which minimizes the $L_1$ norm of the gradient to preserve discontinuities. While effective at mitigating instabilities, these handcrafted priors are rigid and often suppress fine-scale geological features. Recent advances in generative modeling open new possibilities for seismic inversion. Variational autoencoders (VAEs) and generative adversarial networks (GANs) \cite[]{mosser2020stochastic, laloy2018training} have been explored as learned priors, but their ability to capture realistic geological variability can be limited. Diffusion generative models \cite[]{ho2020denoising, song2020score} provide another learned-prior framework by modeling data distributions through iterative denoising of a stochastic process. When trained on representative geological models, they can encode structural patterns that are useful as priors for ill-posed inverse problems.

In geoscience, two main research directions have recently emerged for integrating diffusion models with inversion workflows. The first modifies the diffusion model architecture into multimodal frameworks that jointly learn from velocity models and paired measurements such as seismic waveforms or well log data \cite[]{wang2024controllable, wang2024seisfusion}. These approaches aim to
enable conditional generation directly aligned with physical observations. The second direction retains the standard diffusion training on geological velocity datasets alone, but introduces physical constraints during the sampling stage
to guide the generative process toward data-consistent reconstructions \cite[]{wang2023prior, taufik2024learned, ravasi2025geophysical}. While both directions are promising, most studies to date focus on relatively linear or
simplified inverse problems, and the design of effective algorithms for nonlinear PDE-constrained settings such as FWI has not been systematically addressed. Moreover, although diffusion models have been successfully applied to imaging tasks including denoising \cite[]{chung2022diffusion}, inpainting \cite[]{lugmayr2022repaint}, and enhancement \cite[]{yin2023cle}, these efforts largely concentrate on linear inverse formulations and offer limited insight into their generalization capability in complex velocity models.

In this paper, we evaluate diffusion-guided regularization methods for FWI under a common PDE-constrained inversion setting. The three coupling strategies are adapted from recent diffusion inverse-problem literature: Manifold-Preserving Guided Diffusion (MPGD) \cite[]{he2023manifold}, which injects FWI updates into diffusion inference; Guided Image Synthesis and Editing (SDEdit) \cite[]{meng2021sdedit}, which denoises a perturbed FWI reconstruction; and Split Gibbs Diffusion Sampling (SGDS) \cite[]{vono2019split}, which separates likelihood-driven FWI updates from prior-driven denoising steps. Our contribution is a geophysical benchmark and interpretation of these coupling choices in a nonlinear wave-equation inverse problem. We emphasize practical questions raised by FWI: how the diffusion noise level should be tuned, how a $100\times100$ prior can be deployed on larger benchmarks, how training distribution affects the learned prior, and how data noise changes reconstruction quality. The theoretical role of SGDS is presented through a local splitting and projected-gradient interpretation that clarifies why alternating likelihood and denoising steps can help stabilize the nonlinear FWI update.

\section{Theory}
Full Waveform Inversion (FWI) is a PDE-constrained optimization problem used to estimate the unknown subsurface model \( \mathbf{m} \) by minimizing the discrepancy between synthetic and observed seismic wavefields. Let \(u(x,t;\mathbf{m})\) denote the simulated wavefield and let \(d_r(t)\) denote the observed trace at receiver \(r\). If \(R\) samples the wavefield at the receiver locations, then \((Ru)_r(t)=u(x_r,t;\mathbf{m})\). The standard FWI formulation is:
\begin{align}
\min_{\mathbf{m}} \; \mathcal{J}(u, \mathbf{m}) &= \tfrac12\sum_r\|(Ru)_r(t)-d_r(t)\|_{L^2(0,T)}^2 + \lambda \mathcal{T}(\mathbf{m})\\
&\text{subject to} \quad \mathcal{F}(\mathbf{m}) u = s
\end{align}
Here, \( R \) is the receiver sampling operator, \( \mathcal{F}(\mathbf{m}) \) represents the discretized PDE forward operator (e.g., acoustic or elastic wave equation), and \( \mathcal{T}(\mathbf{m}) \) is a regularization term that encodes prior assumptions about the model \( \mathbf{m} \), with \( \lambda \) as the regularization weight.

Because the inversion is highly nonlinear and ill-posed, particularly when the starting model lacks low-wavenumber accuracy, prior information is essential to stabilize the reconstruction and avoid cycle skipping. Traditionally, such priors are encoded via analytic function space regularization (e.g., Sobolev or TV norms). In this section, we propose an alternative: using data-driven priors learned via diffusion models that constrain the solution to a learned manifold of plausible subsurface structures.

The logic of the section is as follows. We first recall the role of classical regularization in FWI and the type of stability it can provide under standard variational assumptions. We then reinterpret diffusion denoisers as learned regularizers: the denoiser or score complements the wave-equation objective by supplying a data-driven structural prior. Finally, we use a local splitting argument to explain why the SGDS update can behave like a stable projected-gradient scheme near the learned geological model class.
\subsection{FWI formulation}

Regularization stabilizes the nonlinear full waveform inversion (FWI) problem and incorporates prior information about the Earth's subsurface. We present both the classical variational formulation and its data-driven extension based on diffusion generative models. The former constrains the model in a convex functional space, while the latter implicitly projects it onto a nonlinear manifold of geologically plausible structures.

\subsubsection{Classical variational regularization}

In many ill-posed inverse problems, particularly PDE-constrained ones such as Full Waveform Inversion (FWI), it is necessary to introduce prior knowledge about the unknown model \( m \in \mathcal{X} \), where \( \mathcal{X} \) is typically a Banach or Hilbert space such as \( L^2(\Omega) \). Regularization stabilizes the inversion, mitigates noise amplification, and embeds useful structural assumptions into the solution. In the variational setting, FWI seeks
\begin{equation}
m^\star = \arg\min_{m \in \mathcal{X}}
\Big[\tfrac12\|F(m)-d\|^2_{L^2(\Gamma)} + \lambda\,\Phi(m)\Big],
\label{eq:var_fwi}
\end{equation}
where \( F \) is the forward wave operator, \( d \) are the observed data, and \( \Phi \) is a stabilizing regularizer that incorporates prior information about the model.

\textbf{Sobolev Regularization.} A common choice is the Sobolev (Tikhonov-type) regularization, which assumes that the model belongs to a Sobolev space \( H^s(\Omega) \). The corresponding penalty,
\begin{equation}
\Phi_{\text{H}^s}(m) = \| m \|_{H^s}^2 = \sum_{|\alpha|\le s} \| D^\alpha m \|_{L^2(\Omega)}^2,
\end{equation}
enforces smoothness at a prescribed scale and suppresses high-frequency artifacts. Here, \(D^\alpha\) denotes the weak derivative associated with the spatial multi-index \(\alpha\), and \(\Omega\) is the physical model domain. The data norm \(L^2(\Gamma)\) in equation~\eqref{eq:var_fwi} is taken over the acquisition boundary or receiver-time measurement set \(\Gamma\). By tuning the Sobolev order \( s \), one can balance the recovery of large-scale trends against the suppression of fine-scale oscillations. This makes Tikhonov regularization particularly effective for capturing background velocity structures in FWI, where overly detailed fluctuations are not supported by the data.

\textbf{Total variation Regularization.} Alternatively, the Total Variation (TV) functional,
\begin{equation}
\Phi_{\text{TV}}(m) = \|\nabla m\|_1 = \int_\Omega |\nabla m(x)|\,dx,
\end{equation}
promotes sparsity of the gradient and thus favors piecewise-constant reconstructions with sharp discontinuities. This property is especially valuable for seismic applications, where subsurface models often contain blocky structures such as stratified layers, faults, and salt boundaries. Unlike quadratic smoothing penalties, TV regularization better preserves discontinuities and reduces oversmoothing of geologic interfaces, leading to reconstructions that are more geologically plausible.

\textbf{Existence and Stability.} Let \( F : BV(\Omega)\cap L^2(\Omega) \to L^2(\Gamma) \) be continuous and weak* sequentially closed. Then, the classical direct method of the calculus of variations ensures:

\begin{theorem}[Existence of Minimizers]
\label{thm:existence_classical}
The functional~\eqref{eq:var_fwi} with \( \Phi = \mathrm{TV} \) admits at least one minimizer \( m_\lambda \in BV(\Omega) \). Moreover, if the data \( d^\delta = F(m^\dagger) + \eta^\delta \) satisfy \( \|\eta^\delta\|_{L^2} \le \delta \) and \( \lambda(\delta) \to 0 \) with \( \delta^2/\lambda(\delta) \to 0 \), then \( m_{\lambda(\delta)}^\delta \to m^\dagger \) (up to subsequences) in \( L^1(\Omega) \).
\end{theorem}

Hence, classical convex regularization provides well-posedness of the inverse problem in the sense of existence, stability, and convergence as the noise level tends to zero. However, despite their success in stabilizing inversion and enhancing structural interpretability, such handcrafted priors remain limited: they impose explicit functional constraints (e.g., smoothness or piecewise constancy) but fail to capture the richer geometric or statistical structure of realistic geological media. This limitation motivates the development of data-driven regularization strategies, such as diffusion-based learned priors, that can encode complex manifold structures beyond classical function spaces.

\subsubsection{Diffusion models as learnable priors}

Recent advances in generative modeling provide an alternative to explicit functional norms by learning a data-driven prior over the model space. Instead of assuming that the solution lies in a pre-defined function space such as \( H^1(\Omega) \) or \( BV(\Omega) \), we assume it lies near a low-dimensional manifold of geologically plausible structures, inferred directly from training data via diffusion probabilistic models.

\textbf{Bayesian interpretation.} In the Bayesian framework, the posterior distribution of the subsurface model is
\begin{equation}
p(m \mid d) \propto p(d \mid m)\,p(m),
\end{equation}
where the likelihood \(p(d \mid m)\) measures data consistency and the prior \(p(m)\) encodes structural assumptions. Classical regularization corresponds to specifying \(p(m)\) analytically via convex norms, leading to deterministic optimization of the form
\[
\min_m \; \|F(m)-d\|^2 + \lambda \Phi(m).
\]
Diffusion generative models provide an implicit statistical prior by learning the probability distribution of realistic velocity models from data. They learn the score function
\begin{equation}
s_\theta(x) \approx \nabla_x \log p(x),
\label{eq:score_def}
\end{equation}
where \(s_\theta\) is the gradient of the log-density of the learned data distribution. The learned score function can therefore be incorporated directly into the inversion, acting as a nonlinear regularization term that promotes geologically realistic structures:
\begin{equation}
J_\lambda(m) = \tfrac12\|F(m)-d\|^2_{L^2(\Gamma)} - \lambda \log p_\theta(m),
\label{eq:diffusion_reg}
\end{equation}
where \(p_\theta\) is the probability distribution represented implicitly by the trained diffusion model. The learned score \(s_\theta(m) = \nabla_m \log p_\theta(m)\) then acts as a nonlinear regularization gradient guiding the inversion toward the manifold of realistic geological structures.

\textbf{Score-based formulation.} Diffusion training learns a denoiser that approximates the maximum a posteriori (MAP) solution of a noisy input:
\begin{equation}
\mathrm{Denoiser}(x_t)
\approx \arg\max_{x'}\!\left[-\tfrac{1}{2\sigma_t^2}\|x'-x_t\|^2+\log p(x')\right].
\end{equation}
In the small-noise limit, the denoiser residual satisfies
\begin{equation}
\nabla_x \log p(x) \approx \tfrac{1}{\sigma^2}\big(\mathrm{Denoiser}(x)-x\big),
\end{equation}
revealing that the learned denoiser implicitly parameterizes the gradient of the log-prior. Consequently, diffusion inference performs alternating updates between the data likelihood and the denoising prior, analogous to classical alternating projection or proximal steps.

\textbf{Local splitting interpretation.} We use a local model to connect SGDS with learned regularization in the nonlinear FWI setting. Let $P_\theta$ denote the denoising map induced by the pretrained diffusion model, and let $J(m)$ be the FWI data-misfit objective used in the current frequency band. A simplified SGDS step can be written as
\begin{equation}
T_\alpha(m)=P_\theta\big(m-\alpha\nabla J(m)\big),
\label{eq:proj_iter}
\end{equation}
which is a projected-gradient-like update in which the physics step reduces data misfit and the denoising step pulls the iterate toward the learned geological model class $\Sigma_\theta$. In this view, SGDS is a controlled learned regularization scheme rather than a purely black-box denoising heuristic.

This local interpretation can be made more explicit. Let $\Sigma_\theta$ denote the learned geological model class and suppose that, in a neighborhood of a reference point $m^\star\in\Sigma_\theta$, the denoiser $P_\theta$ behaves as an approximate projection onto $\Sigma_\theta$ with projection defect $\delta$. If $\nabla J$ is locally Lipschitz and $J$ is locally coercive along tangent directions of $\Sigma_\theta$, then a standard projected-gradient expansion gives
\begin{equation}
\|T_\alpha(m)-m^\star\|
\le
C_T\|e_T\|
+ C_{\rm man}\operatorname{dist}(m,\Sigma_\theta)
+ C_{\rm proj}\delta\|m-m^\star\|
+ O(\|m-m^\star\|^2),
\label{eq:sgds_local_stability}
\end{equation}
where $e_T$ is the tangent component of the local error $m-m^\star$. When the stepsize is chosen so that $C_T<1$, the SGDS map is contractive along tangent directions, up to the deviation from the learned manifold and the denoiser projection error. This local bound provides an algorithmic interpretation of the observed alternating behavior: a data-consistency step reduces the FWI misfit, and a denoising-prior step returns the iterate toward the structural support represented in the training data.

\begin{table}[h!]
\centering
\caption{Interpretation of classical and diffusion-based regularization used in this work. The diffusion entry is a local algorithmic interpretation rather than a global convergence theorem.}
\label{tab:reg_compare}
\resizebox{\textwidth}{!}{%
\begin{tabular}{lll}
\toprule
\textbf{Property} & \textbf{Classical (Tikhonov/TV)} & \textbf{Diffusion prior} \\
\midrule
Geometry & Convex function space ($H^1$, $BV$) & Learned geological model class \\
Update & Analytical proximal/gradient step & Denoising map $P_\theta$ \\
Role & Smoothness or edge preservation & Data-driven structural regularization \\
Guarantee & Classical stability under assumptions & Local stability interpretation under stated assumptions \\
Main risk & Oversmoothing or blocky bias & Training-distribution bias \\
\bottomrule
\end{tabular}%
}
\end{table}

\section{Diffusion model training and sampling}
The previous section treated the learned prior abstractly through a score field or denoising map \(P_\theta\). This section specifies the practical realization used in the experiments: a DDPM denoiser trained on GeoFWI velocity patches and then coupled to FWI through Tweedie's estimate and measurement-guided updates.
\subsection{Score-based diffusion framework}
Score-based diffusion models (SGMs) \cite{song2019generative,song2020score} learn a score field \( \nabla_x \log p_t(x) \) that points noisy samples back toward the data distribution. This makes them useful as learned priors in inverse problems: the score or denoiser can encourage iterates to remain close to the distribution represented by the training velocity models. We briefly summarize the formulation used in this work to fix notation.

The forward process gradually corrupts clean data \( x_0 \sim p(x) \) into noise \( x_1 \sim \pi(x) \) over a continuous time horizon \( t \in [0,1] \), where \( \pi(\mathbf{x}) \) is a pre-defined noise distribution such as \( \mathcal{N}(0, I)\). This evolution is modeled by the Ito SDE
\begin{equation}
dx_t = f(x_t, t)\,dt + g(t)\,d\omega_t ,
\end{equation}
where \( f(x_t, t) \) is the drift coefficient, \( g(t) \) is the diffusion coefficient, and \( \omega_t \) is Brownian motion. The reverse process uses the learned score to remove noise:
\begin{equation}
dx_t = \Big[f(x_t,t) - g(t)^2 \nabla_{x_t} \log p_t(x_t)\Big]dt + g(t)\,d\omega_t .
\end{equation}
Since the true score is unknown, it is approximated by a neural network \( s_\theta(x_t,t) \) trained by denoising score matching,
\begin{equation}
\theta^* = \arg \min_{\theta} \; \mathbb{E}_{t \sim \mathbb{U}[0,1],\, x_0 \sim p_{\text{data}},\, x_t \sim p(x_t|x_0)} 
\Big[ \| s_\theta(x_t, t) - \nabla_{x_t} \log p(x_t | x_0) \|_2^2 \Big].
\end{equation}
For FWI, the important point is not the generative model by itself, but how the resulting denoiser is coupled to a wave-equation likelihood update.

\subsection{DDPM Specialization}

In practice, we adopt the Denoising Diffusion Probabilistic Model (DDPM) \cite{ho2020denoising}, a discrete-time implementation of the above framework. The Gaussian forward process is
\[
q(x_t \mid x_{t-1}) = \mathcal{N}\!\left(\sqrt{1-\beta_t}\,x_{t-1}, \, \beta_t I\right),
\]
where \(\beta_t\) is the noise schedule. Training is performed by predicting the Gaussian noise added at each step:
\begin{equation}
\theta^* = \arg \min_\theta \; \mathbb{E}_{x_0,\epsilon,t} \Big[ \|\epsilon - \epsilon_\theta(x_t, t)\|_2^2 \Big],
\end{equation}
with \( x_t = \sqrt{\bar{\alpha}_t}\,x_0 + \sqrt{1-\bar{\alpha}_t}\,\epsilon \), \( \epsilon \sim \mathcal{N}(0, I) \), and \( \bar{\alpha}_t = \prod_{j=1}^{t} (1 - \beta_j) \). All results in this work use a DDPM trained on geophysical velocity-model patches.

\subsection{Conditional diffusion guided by FWI}

In many inverse problems, it is challenging to express the conditional distribution \( p(x \mid y) \) directly or to construct explicit conditional samplers. Within the SGM framework, this can be addressed by modifying the reverse diffusion dynamics using Bayes' theorem \cite{chung2022diffusion}. Specifically, the conditional score decomposes as
\begin{equation}
\nabla_{x_t} \log p_t(x_t \mid y) 
= \nabla_{x_t} \log p_t(x_t) + \nabla_{x_t} \log p_t(y \mid x_t).
\end{equation}

The first term is the unconditional score, approximated by the trained network \( \mathbf{s}_{\theta^*}(x_t, t) \).  
The second term incorporates the measurement \( y \) but is generally intractable at intermediate states \( x_t \). To make progress, we leverage Tweedie's formula \cite{efron2011tweedie} to approximate the posterior expectation
\( \hat{x}_0 = \mathbb{E}[x_0 \mid x_t]\),
which provides a denoised estimate of the clean sample given the noisy state.  
For DDPM, Tweedie's estimate is
\begin{equation}
\hat{x}_0 = \frac{1}{\sqrt{\bar{\alpha}_t}} 
\Big( x_t + (1-\bar{\alpha}_t)\,\nabla_{x_t}\log p_t(x_t) \Big).
\end{equation}
Replacing the exact score with the trained network gives
\begin{equation}
\hat{x}_0 \approx \frac{1}{\sqrt{\bar{\alpha}_t}}
\Big( x_t + (1-\bar{\alpha}_t)\,\mathbf{s}_{\theta^*}(x_t, t) \Big).
\end{equation}

The conditional likelihood term is then approximated as
\begin{equation}
\nabla_{x_t} \log p_t(y \mid x_t) 
\approx - \nabla_{x_t} \| y - Ru(\hat{x}_0(x_t)) \|_2^2,
\end{equation}
where \(u\) is the wavefield governed by the acoustic wave equation, \(R\) extracts the boundary measurements, and \(y\) is the observed receiver data. Using the adjoint-state method, this gradient is computed as
\begin{equation}
\nabla_{x_t} \| y - R u(\hat{x}_0(x_t)) \|_2^2
= - \sum_i \int_0^{\tilde{T}} \frac{\partial^2 u_i}{\partial \tilde{t}^2} \, w_i \, d\tilde{t},
\end{equation}
where \(w_i\) is the adjoint wavefield, \(\tilde{t}\) the acoustic time variable, and \(\hat{x}_0\) the estimated velocity model. Combining both contributions, the conditional score used in the reverse SDE is approximated as
\begin{equation}
\nabla_{x_t} \log p_t(x_t \mid y) 
\;\approx\; \mathbf{s}_{\theta^*}(x_t, t) 
- \rho \,\nabla_{x_t} \| y - R u(\hat{x}_0(x_t)) \|_2 ,
\end{equation}
where \( \rho \) is a step-size parameter balancing the learned prior and the data consistency term.  
Thus, the score network \( \mathbf{s}_{\theta^*} \) enforces plausibility with respect to the learned prior, while the PDE-based gradient enforces consistency with seismic observations.

\section{Diffusion-guided inversion algorithms}
We consider three diffusion-guided inversion strategies that incorporate physical gradients and learned priors in different ways. MPGD and SDEdit were originally developed for measurement-guided image generation and editing problems, such as image restoration, super-resolution, or inpainting, where the forward operator is often simpler than a wave-equation modeling loop. FWI is different: each likelihood update requires PDE solves, the objective is strongly nonlinear, and the gradient can be misleading when the starting model is poor. We therefore use MPGD and SDEdit as direct diffusion-guidance baselines, and implement SGDS as an FWI-specific split scheme that separates deterministic wave-equation likelihood updates from diffusion-prior denoising.

In this comparison, MPGD integrates FWI optimization directly into each diffusion step, SDEdit initializes diffusion from an intermediate noisy version of an FWI solution, and SGDS alternates explicitly between likelihood-driven inversion and prior-driven diffusion updates. These approaches can be interpreted through an augmented posterior perspective, but are implemented here as practical diffusion-guided optimization schemes for FWI rather than exact posterior samplers. To ensure clarity and consistency, we summarize the notation used throughout this section:

\begin{itemize}
    \item $x_t$: latent variable at diffusion step $t$, with $t = T, \dots, 0$.
    \item $\hat{x}_0$: denoised estimate of the clean model at each step.
    \item $\epsilon_t \sim \mathcal{N}(0,I)$: Gaussian noise at step $t$.
    \item $\mathbf{s}_\theta(x_t,t)$: score network approximating $\nabla_{x_t} \log p_t(x_t)$.
    \item $\bar{\alpha}_t$: cumulative product of noise schedule coefficients, $\bar{\alpha}_t = \prod_{j=1}^t (1-\beta_j)$.
    \item $A$: forward operator for FWI, mapping model parameters to simulated data.
    \item $y$: observed seismic data at receiver locations.
    \item $K$: number of outer iterations in SGDS.
\end{itemize}

\subsection{Manifold-Preserving Guided Diffusion (MPGD)}

\begin{figure}[H]
\centering
\includegraphics[width=\columnwidth]{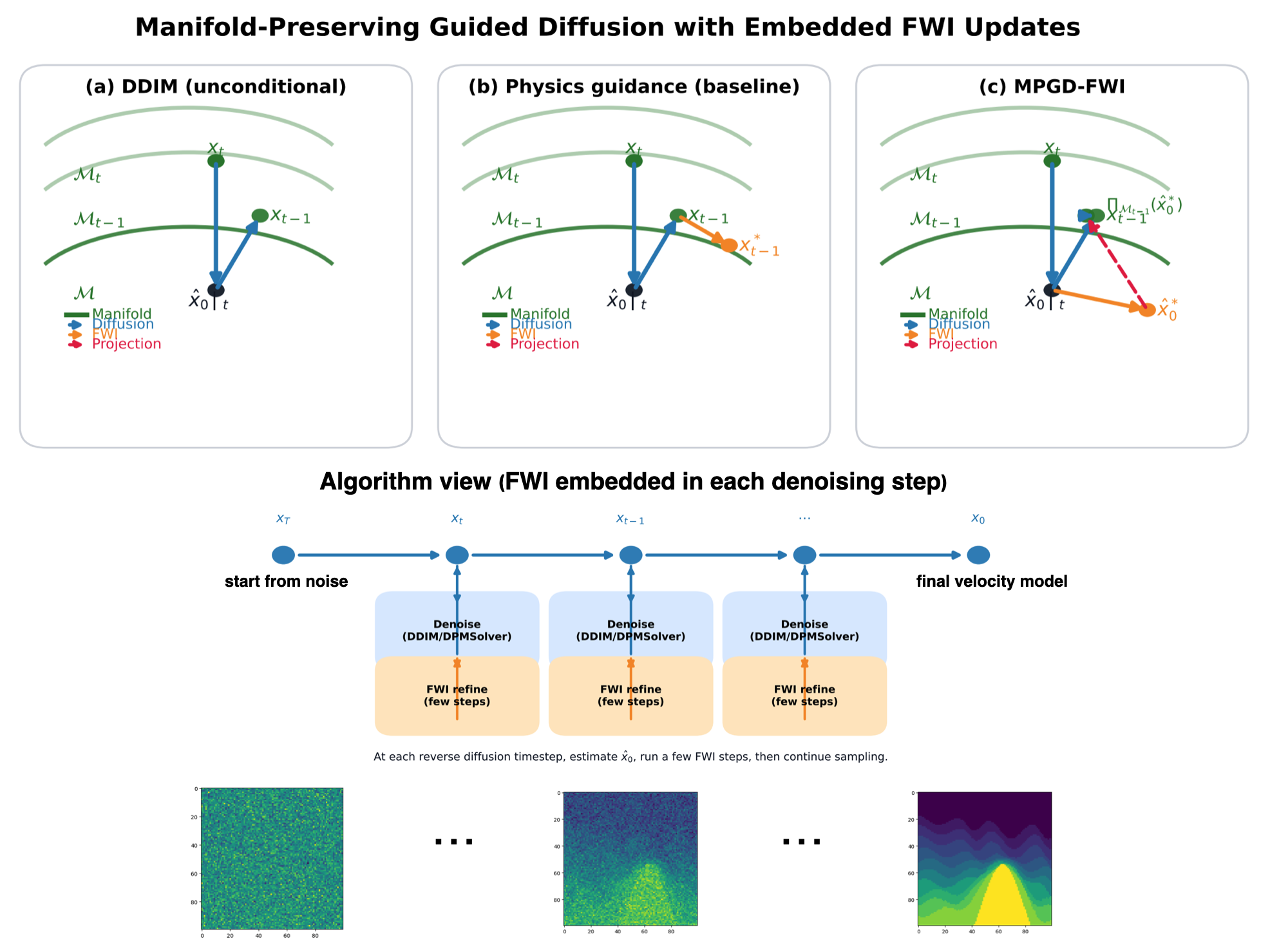}
\caption{Illustration of the Manifold-Preserving Guided Diffusion (MPGD) method, where FWI iteratively refines \( \hat{x_0}\) during diffusion inference to integrate physical constraints while preserving the learned generative manifold.}
\label{fig:MPGD}
\end{figure}

The Manifold-Preserving Guided Diffusion (MPGD) method \cite{he2023manifold} incorporates gradient information from FWI into the diffusion inference process while preserving the learned diffusion manifold. In MPGD, when estimating $\hat{x}_0$ at each step of diffusion inference using Tweedie's formula, the gradient obtained from FWI is first used to update $\hat{x}_0$ through iterative optimization (e.g., with L-BFGS). The updated $\hat{x}_0$ is then reintroduced into the diffusion process for further denoising. This approach combines diffusion inference and FWI iterations, where the prior knowledge from diffusion is fused with physical constraints from FWI. The point at which FWI iterations begin can be adjusted based on problem complexity and reconstruction requirements (Figure~\ref{fig:MPGD}).

\begin{algorithm}[H]
\caption{Manifold-Preserving Guided Diffusion (MPGD)}
\begin{algorithmic}[1]
\State Initialize $x_T \sim \mathcal{N}(0,I)$
\For{$t = T, T-1, \dots, 1$}
    \State Estimate $\hat{x}_0 = \tfrac{1}{\sqrt{\bar{\alpha}_t}}\big(x_t + (1-\bar{\alpha}_t)\mathbf{s}_\theta(x_t,t)\big)$
    \State Refine $\hat{x}_0$ with a few FWI optimization steps (e.g., L-BFGS)
    \State Sample $\epsilon_t \sim \mathcal{N}(0,I)$
    \State Update $x_{t-1} = \sqrt{\bar{\alpha}_{t-1}}\hat{x}_0 + \sqrt{1-\bar{\alpha}_{t-1}}\epsilon_t$
\EndFor
\State \Return $x_0$
\end{algorithmic}
\end{algorithm}

\subsection{SDEdit-guided inversion}

\begin{figure}[H]
\centering
\includegraphics[width=0.9\columnwidth]{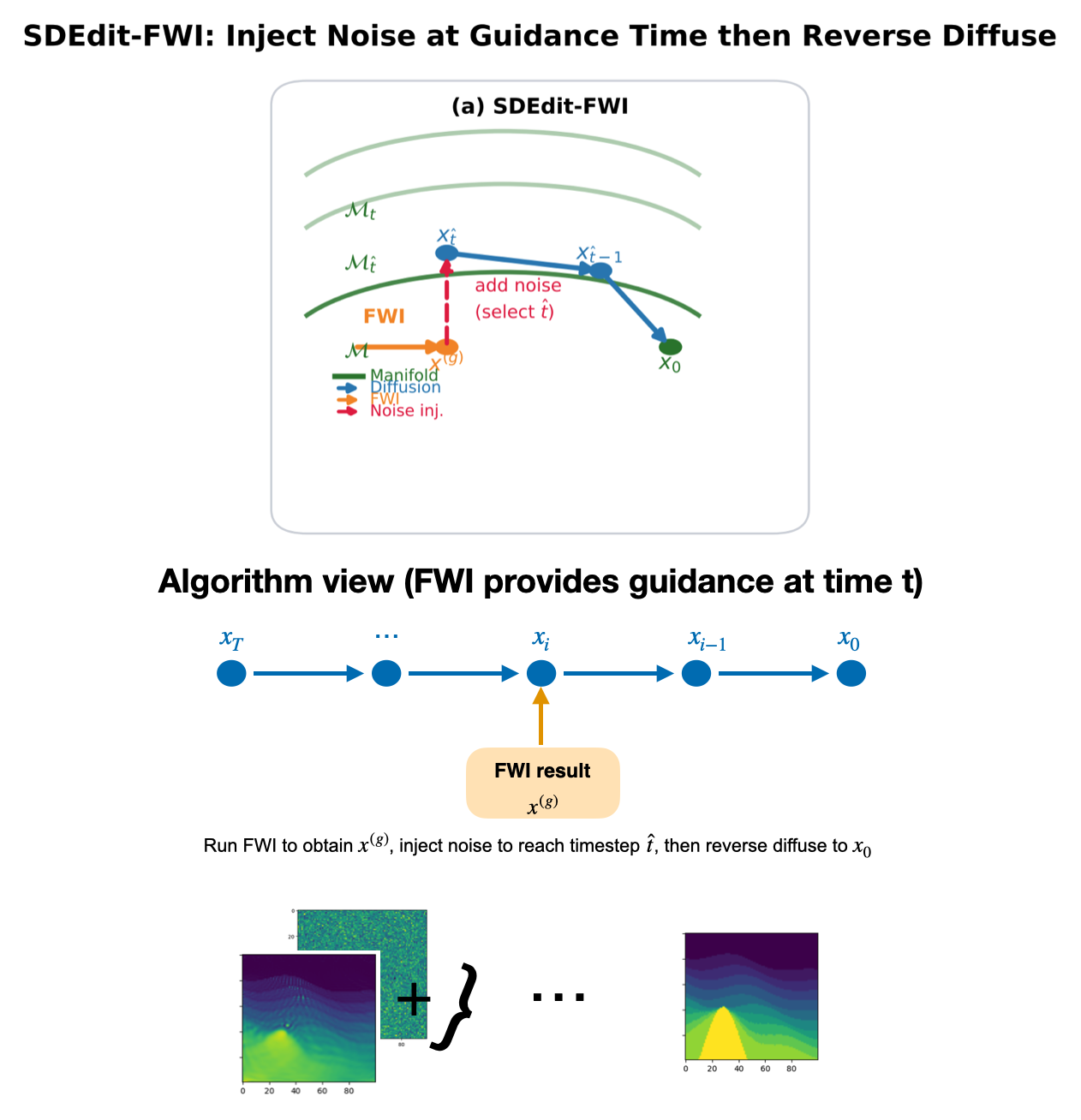}
\caption{SDEdit process with FWI result as guidance at time \( \hat{t} \) for reverse diffusion reconstruction.}
\label{fig:SDEdit}
\end{figure}

The SDEdit method \cite{meng2021sdedit} extends the conventional reverse SDE framework by allowing the generation process to start not only from $t_0=1$ but from any intermediate time $\hat{t} \in [0,1]$. This enables more flexible guidance, as we can inject an FWI result at $\hat{t}$ and then denoise backward to $t=0$. A larger $\hat{t}$ introduces more noise (encouraging realism but potentially deviating from the guide), while a smaller $\hat{t}$ retains more 
information from the FWI initialization but may limit the ability to generate natural-looking results. Thus, a suitable $\hat{t}$ balances realism and data consistency (Figure~\ref{fig:SDEdit}).

\begin{algorithm}[H]
\caption{Guided Image Synthesis and Editing (SDEdit)}
\begin{algorithmic}[1]
\State Given FWI result $x^{(g)}$
\State Choose guidance time $\hat{t}$
\State Sample $x_{\hat{t}} \sim \mathcal{N}(x^{(g)}, \sigma^2(\hat{t})I)$
\For{$t = \hat{t}, \hat{t}-1, \dots, 1$}
    \State Estimate $\hat{x}_0 = \tfrac{1}{\sqrt{\bar{\alpha}_t}}\big(x_t + (1-\bar{\alpha}_t)\mathbf{s}_\theta(x_t,t)\big)$
    \State Sample $\epsilon_t \sim \mathcal{N}(0,I)$
    \State Update $x_{t-1} = \sqrt{\bar{\alpha}_{t-1}}\hat{x}_0 + \sqrt{1-\bar{\alpha}_{t-1}}\epsilon_t$
\EndFor
\State \Return $x_0$
\end{algorithmic}
\end{algorithm}

\subsection{Split Gibbs Diffusion Sampling (SGDS)}

\begin{figure}[H]
\centering
\includegraphics[width=\columnwidth]{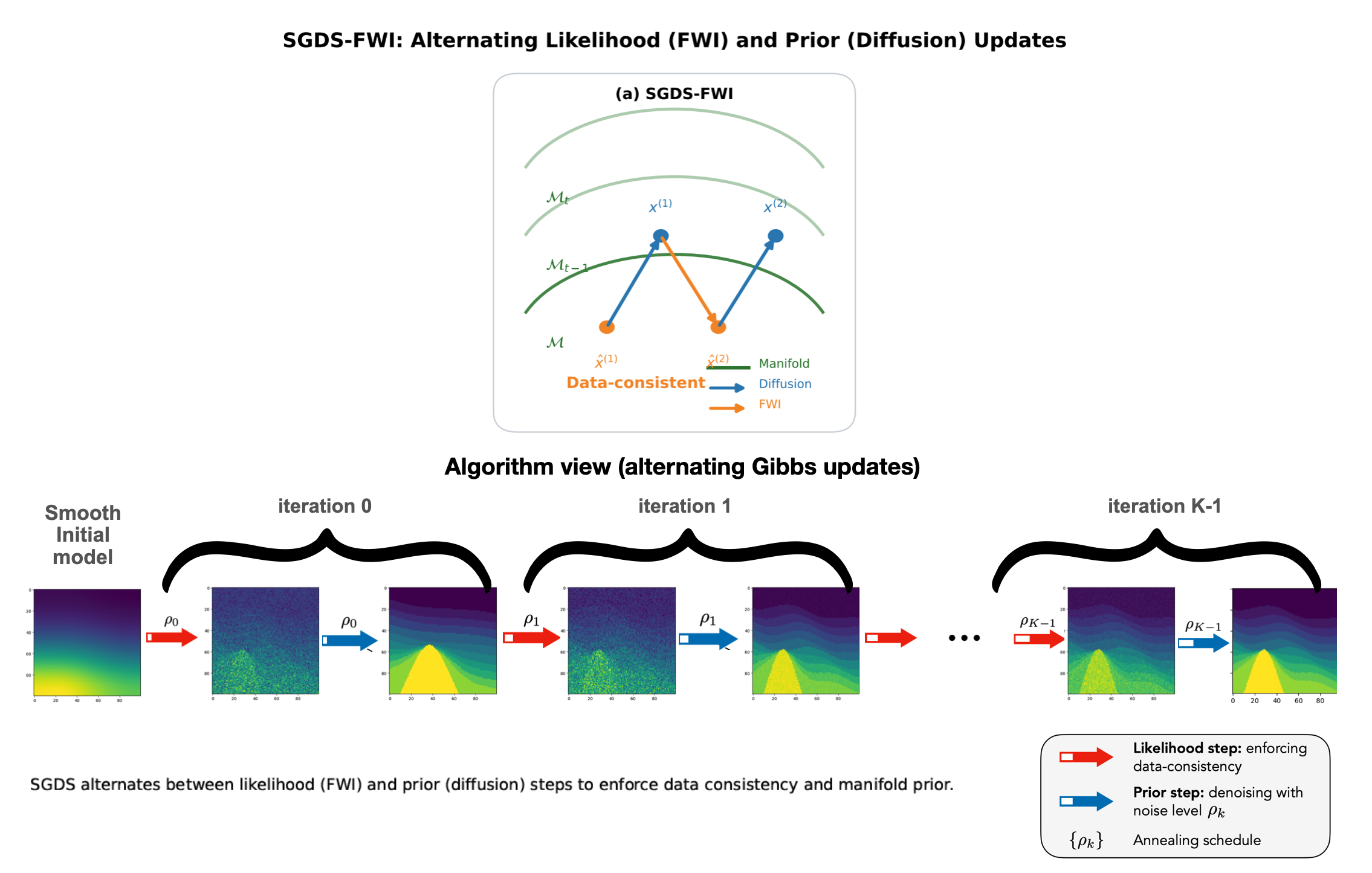}
\caption{Workflow of SGDS Method, where FWI and diffusion processes alternate iteratively, refining the velocity model through separate likelihood and prior inference cycles.}
\label{fig:SGDS}
\end{figure}

The Split Gibbs Diffusion Sampling (SGDS) method is motivated by Bayesian inference via variable splitting. Formally, one can write a posterior density $p(x|y) \propto \exp\big(-f(x;y) - g(x)\big)$, where $f(x;y) := -\log p(y|x)$ is the likelihood potential defined by the data misfit, and $g(x) := -\log p(x)$ is the prior potential. Introducing an auxiliary variable $z$ gives the augmented joint density $\pi(x,z) \propto \exp\!\left(-f(z;y) - g(x) - \frac{1}{2\rho^2}\|x-z\|_2^2\right)$, where $\rho$ is a coupling strength (Figure~\ref{fig:SGDS}). In this paper, we use this split as an optimization-inspired alternating scheme rather than as an exact posterior sampler:

\begin{itemize}
    \item \textbf{Likelihood step:} update $z^{(k)}$ given $x^{(k)}$,
    \[
    z^{(k)} \sim \pi^{Z|X=x^{(k)}}(z) \propto 
    \exp\!\left(-f(z;y) - \frac{1}{2\rho^2}\|x^{(k)}-z\|_2^2\right).
    \]
    \item \textbf{Prior step:} update $x^{(k+1)}$ given $z^{(k)}$,
    \[
    x^{(k+1)} \sim \pi^{X|Z=z^{(k)}}(x) \propto 
    \exp\!\left(-g(x) - \frac{1}{2\rho^2}\|x-z^{(k)}\|_2^2\right).
    \]
\end{itemize}

In practice, these two steps can be directly mapped to the FWI + diffusion framework. The likelihood step corresponds to updating the model with FWI, regularized towards the previous iterate. The prior step corresponds to running a reverse diffusion process that incorporates the learned generative prior. The
resulting iterative procedure is summarized below. In our implementation, the coupling strength is controlled implicitly through the noise level $\sigma_k$ used to perturb the iterate before denoising.

\begin{algorithm}[H]
\caption{Split Gibbs Diffusion Scheme (SGDS) for FWI with Diffusion Prior}
\label{alg:sgds}
\begin{algorithmic}[1]
\Require \text{Observed seismic data $\mathbf{y}$, forward operator $\mathcal{A}(\cdot)$, pretrained score model}
\State $\mathbf{x}^{(0)} \leftarrow \arg\min_{\mathbf{x}} \|\mathbf{y} - \mathcal{A}(\mathbf{x})\|_2^2$ \hfill $\triangleright$ \text{FWI initialization}
\For{$k = 1,2,\dots,K$}
    \State \textbf{Likelihood step (FWI update):} \\
    Solve a penalized least-squares subproblem to enforce data consistency:
    \[
      \hat{\mathbf{x}}^{(k)} 
      = \arg\min_{\mathbf{x}} \; \| \mathbf{y} - \mathcal{A}(\mathbf{x}) \|_2^2 
        + \tfrac{1}{2\sigma_k^2}\| \mathbf{x} - \mathbf{x}^{(k-1)} \|_2^2,
    \]
    which acts as a proximal gradient step on the misfit.
    
    \State \textbf{Stochastic perturbation:}
    \[
    \mathbf{z}^{(k)} \leftarrow \hat{\mathbf{x}}^{(k)} + \alpha \cdot \frac{\sigma_k}{\sigma_1} \cdot \boldsymbol{\epsilon}, \quad \boldsymbol{\epsilon} \sim \mathcal{N}(\mathbf{0}, \mathbf{I})
    \]
    
    \State \textbf{Prior step (Diffusion update):} \\
    Refine $\hat{\mathbf{x}}^{(k)}$ using a reverse-diffusion step guided by the learned prior:
    \[
      \mathbf{x}^{(k)} \;=\; \mathcal{D}_\theta(\mathbf{z}^{(k)};\sigma_k),
    \]
    where $\mathcal{D}_\theta$ denotes the denoiser / diffusion operator.
\EndFor
\State \textbf{Output:} $\mathbf{x}^{(K)}$ as the recovered model \hfill $\triangleright$ \text{Final FWI refinement}
\end{algorithmic}
\vspace{0.5em}
\noindent where $\mathcal{D}_\theta$ denotes the single-step Tweedie estimator:
\begin{equation}
    \mathcal{D}_\theta(\mathbf{x};\, \sigma) = \mathbb{E}[\mathbf{x}_0 \mid \mathbf{x}_\sigma = \sqrt{\bar{\alpha}_\sigma}\, \mathbf{x} + \sqrt{1 - \bar{\alpha}_\sigma}\, \boldsymbol{\epsilon}] = \frac{\mathbf{x}_\sigma - \sqrt{1 - \bar{\alpha}_\sigma}\, \boldsymbol{\epsilon}_\theta(\mathbf{x}_\sigma, \sigma)}{\sqrt{\bar{\alpha}_\sigma}}
\end{equation}
\end{algorithm}

This formulation highlights SGDS as a cyclic refinement method: each iteration
balances data consistency (through FWI) and prior regularization (through
diffusion). By gradually adjusting the schedule of the diffusion starting time
$\hat{t}$, SGDS transitions smoothly toward convergence, reducing noise
perturbations in later cycles while preserving exploration in early iterations.

Table~\ref{tab:comparison} highlights the trade-offs among the three approaches. MPGD applies strong physical guidance but can be sensitive to inaccurate FWI gradients. SDEdit is simple and flexible, but its output remains strongly tied to the quality of the initial FWI model and the chosen start time. SGDS is more expensive because it alternates several FWI updates with diffusion-prior steps, but the separation of the likelihood and prior updates makes the method easier to tune in the nonlinear FWI setting.

\begin{table}[t]
\centering
\caption{Practical comparison of the three measurement-guided diffusion methods.}
\label{tab:comparison}
\resizebox{\textwidth}{!}{%
\begin{tabular}{llll}
\toprule
\textbf{Method} & \textbf{Key tuning parameter} & \textbf{Main failure mode} & \textbf{Cost} \\
\midrule
MPGD & FWI strength per reverse step & Biased gradients can corrupt sampling & High \\
SDEdit & Start time $\hat{t}$ & Too small: little repair; too large: loss of data fit & Medium \\
SGDS & Denoising level / coupling schedule & Too weak: no prior repair; too strong: hallucination & Highest \\
\bottomrule
\end{tabular}%
}
\end{table}

\section{Numerical Examples}
To evaluate the effectiveness of diffusion-guided regularization in seismic inversion, we conduct comprehensive numerical experiments using synthetic datasets and benchmark velocity models. This section is divided into five parts: (1) training dataset and model settings, (2) controlled in-distribution GeoFWI tests using three training-free guidance strategies, (3) benchmark-scale transfer tests on Marmousi and Overthrust, (4) sensitivity studies for diffusion level and additive noise, and (5) a more difficult Sigsbee2A salt benchmark used as a poor-starting-model stress test.

\subsection{Training dataset and model configuration}

\begin{figure}[H]
\centering
\includegraphics[width=\columnwidth]{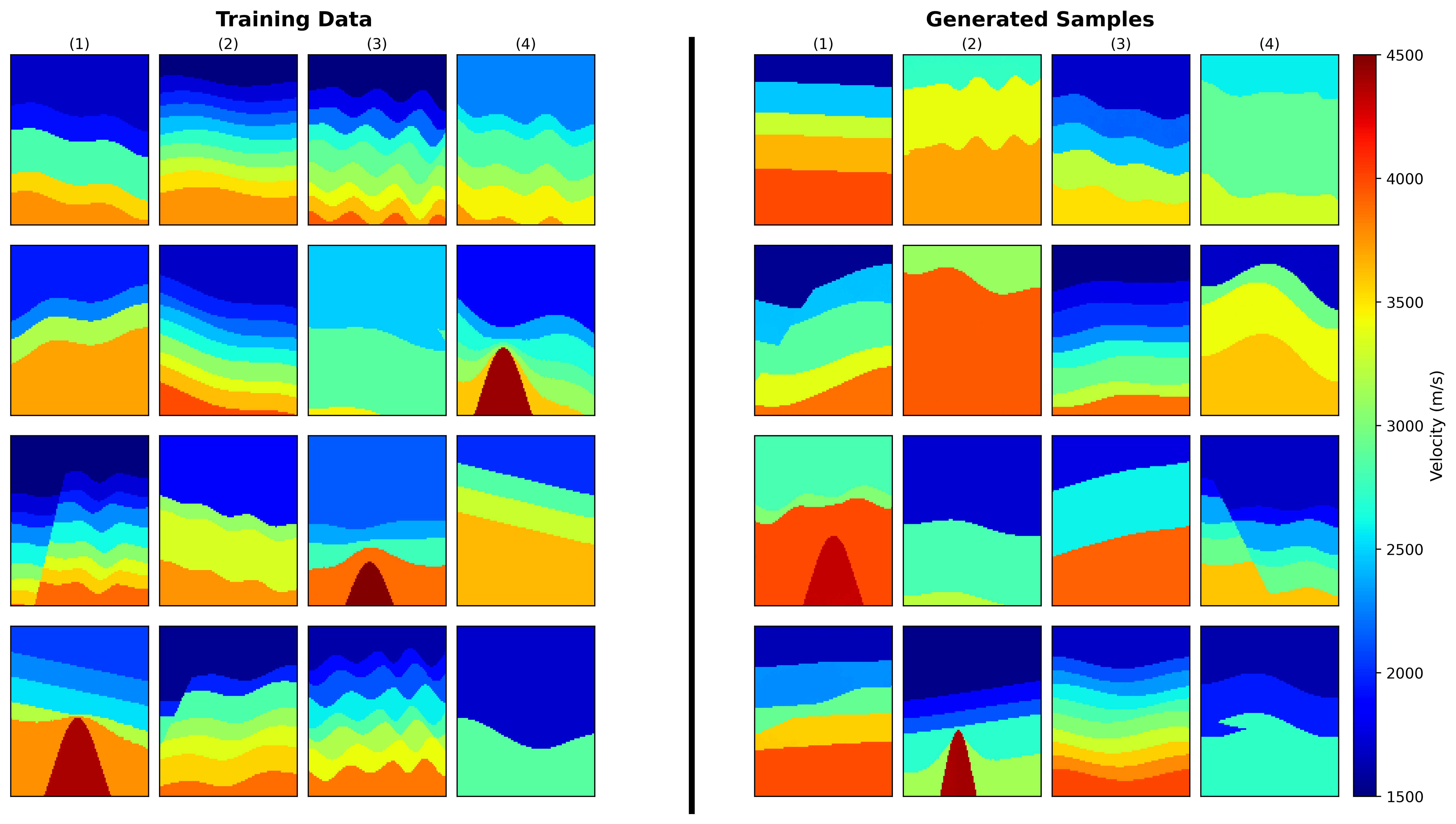}
\caption{Representative training velocity models (left) and diffusion-generated samples (right). The generated models reproduce several structural motifs and velocity contrasts present in the training set, suggesting that the diffusion model has learned a useful synthetic geological prior for guiding inversion under distributional match.
}
\label{fig:train_vs_gen_16}
\end{figure}

The unguided diffusion generative model used in this work figure~\ref{fig:train_vs_gen_16} was trained offline and independently of the downstream inverse problem. Training was performed on a large-scale synthetic dataset representative of plausible subsurface geology, with the following configuration:

\begin{itemize}
    \item \textbf{Training Dataset (GeoFWI):}  
    The dataset contains 45,000 2D velocity models \cite[]{liubin2021geo,ren2021building} of size $100 \times 100$ with a spatial resolution of 10\,m/pixel. These models exhibit diverse geological structures, including salt bodies, dipping faults, and complex stratified layers, with velocity values ranging from 1500\,m/s to 4500\,m/s. Seismic wavefields were simulated using a 25\,Hz Ricker wavelet with 10 sources and 100 receivers, recorded for 4.0s. The initial models were smoothed versions of the ground truth.
    
    \item \textbf{Diffusion Model Architecture:}  
    A U-Net-like denoiser was employed with base dimension 64 and channel multipliers $(1, 2, 8)$ across three levels, resulting in encoder-decoder features of 64--128--512 channels. Each level contains group normalization, SiLU activations, and convolutional blocks. No attention mechanism was used and flash attention was disabled.
    
    \item \textbf{Diffusion and Sampling Settings:}  
    The forward diffusion process consists of 1000 steps with Gaussian noise. Inference was performed with 250 denoising steps using DDIM for accelerated sampling.
    
    \item \textbf{Training Configuration:}  
    Training was performed for 300{,}000 iterations using the Adam optimizer with a learning rate of $8 \times 10^{-5}$, a batch size of 32, and gradient accumulation every 2 steps. Mixed precision (AMP) was enabled. The model has approximately 26 million trainable parameters.
    
    \item \textbf{Implementation Notes:}  
    The model was implemented in PyTorch and trained using distributed data parallelism. No conditioning or guidance was applied during training, making the model suitable for flexible downstream use.
\end{itemize}

\begin{table}[t]
\centering
\caption{Shared benchmark-scale settings used in the Marmousi and Overthrust transfer experiments.}
\label{tab:benchmark_protocol}
\resizebox{\textwidth}{!}{%
\begin{tabular}{ll}
\toprule
\textbf{Category} & \textbf{Configuration} \\
\midrule
Training prior & GeoFWI velocity patches of size $100\times100$ \\
Marmousi geometry & 20 shots in a mixed transmission/reflection setting \\
Overthrust geometry & 40 shots with a full surface receiver array \\
Frequency continuation & Multiscale low-frequency continuation followed by FWI refinement \\
Classical baselines & $L_2$ FWI and $L_2$ FWI with TV regularization \\
TV weight & Uniform TV weight $\lambda_{TV}=0.01$ \\
SGDS outer blocks & Alternating deterministic FWI updates and diffusion denoising \\
Benchmark deployment & Direct patch, downsample-single, downsample-multi, or columnwise denoising \\
\bottomrule
\end{tabular}%
}
\end{table}

Unless otherwise stated, all quantitative reconstruction metrics are computed on the physical velocity model in m/s using the same spatial grid as the reference model. PSNR and SSIM use the reference-model velocity range, $\max(m_{\mathrm{true}})-\min(m_{\mathrm{true}})$, as the data range. RMSE is reported in m/s, NRMSE is RMSE divided by the same reference velocity range, and relative $\ell_2$ error is reported as $\|m-m_{\mathrm{true}}\|_2/\|m_{\mathrm{true}}\|_2$. For each visual comparison, panels within the same experiment use a shared color scale to avoid contrast-induced visual bias.

\begin{table}[t]
\centering
\caption{Measured computational cost on two representative GeoFWI cases using one NVIDIA RTX A4500 GPU per run. SDEdit and SGDS are reported as end-to-end workflows including the common FWI+TV seed. One FWI closure is counted as one forward/adjoint pair per shot. MPGD wall time is reported, but its internal PDE-call count was not separately instrumented.}
\label{tab:cost_accounting}
\resizebox{\textwidth}{!}{%
\begin{tabular}{lrrrrr}
\toprule
\textbf{Method} & \textbf{Layer time (s)} & \textbf{Salt time (s)} & \textbf{GPU-hours avg.} & \textbf{PDE solves} & \textbf{Diffusion calls} \\
\midrule
FWI $L_2$ & 67.9 & 65.7 & 0.0186 & 17.3k & 0 \\
FWI $L_2$+TV & 67.1 & 65.9 & 0.0185 & 17.0k & 0 \\
MPGD & 883.8 & 895.3 & 0.2471 & n/a & 1 \\
SDEdit & 74.4 & 73.2 & 0.0205 & 17.0k & 1 \\
SGDS & 278.8 & 270.0 & 0.0762 & 60.3k & 6 \\
\bottomrule
\end{tabular}%
}
\end{table}

The measured cost confirms the expected trade-off. SDEdit adds only one diffusion call after the FWI+TV seed, so its end-to-end cost is close to the classical baseline. SGDS is more expensive because it performs several additional FWI blocks and denoising steps; in these two cases, its end-to-end wall time is about 4.1 times that of $L_2$ FWI. However, SGDS remains substantially cheaper than MPGD in this implementation, because MPGD embeds the physical guidance more tightly inside diffusion inference. These measurements quantify the computational cost associated with the reconstruction-quality and robustness gains reported for SGDS.

The training distribution controls the geological vocabulary of the learned prior. Under distributional match, the GeoFWI experiments show that the coupling mechanism can transfer this vocabulary into the FWI update. In additional dissertation experiments, enriching the salt subset broadened the generated salt morphologies and improved salt recovery, confirming that the learned prior's geological vocabulary directly affects inversion quality.

\begin{figure}[H]
\centering
\includegraphics[width=\columnwidth]{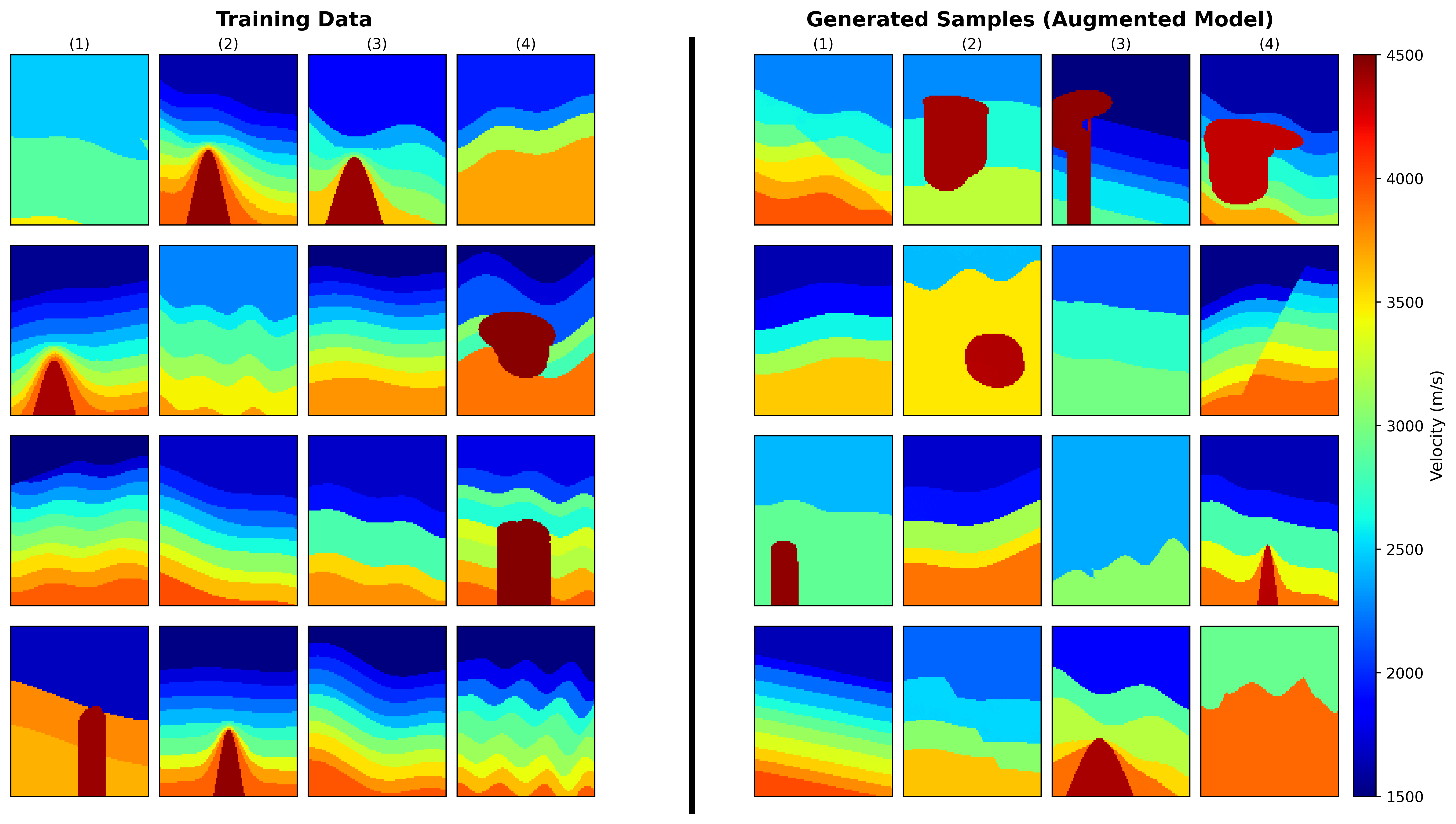}
\caption{Representative augmented salt training models (left block) and unconditional diffusion-generated samples after salt enrichment (right block). The enriched training set broadens the learned salt vocabulary, enabling the diffusion prior to repair a wider range of salt geometries during inversion.}
\label{fig:train_vs_gen_4x4_aug}
\end{figure}

\subsection{Synthetic geological models}

To evaluate the effectiveness of different training-free guidance algorithms in incorporating physical gradient information into the diffusion generation process, we conduct controlled experiments on three representative geological scenarios: salt body, fault, and complex layered structures. These examples are held-out GeoFWI models drawn from the same structural family as the training data, providing controlled in-distribution tests of the proposed coupling mechanism. The visual comparisons and diffusion-level sweep use representative diagnostic cases; to reduce sensitivity to case selection, we additionally ran two newly selected held-out examples per category and report mean and standard deviation over 18 complete GeoFWI inversions in Table~\ref{tab:geofwi_heldout_stats}.

We compare three conditional guidance strategies: Manifold-Preserving Guided Diffusion (MPGD), SDEdit, and Split Gibbs Diffusion Sampling (SGDS). Each method operates on the same initial inversion output and applies guidance via physical gradients computed from the full waveform inversion (FWI) objective. Classical inversion baselines, including standard FWI and FWI with total variation (TV) regularization, are included to benchmark the performance of learned priors.

This setup allows us to investigate two central questions: 
\begin{enumerate}
    \item How effective is gradient-based guidance in nonlinear physical models for directing generative sampling?
    \item Can diffusion-based regularization outperform handcrafted priors such as TV in terms of structural fidelity and resolution?
\end{enumerate}

The salt structure presents a particularly challenging case. In this scenario, the FWI gradient signal is concentrated primarily along the salt top boundary and tends to vanish within the salt body, making it difficult to reconstruct the full geometry using traditional methods. In contrast, our guided diffusion approaches are able to balance the influence of the physical gradient and the learned prior to better recover the salt body while preserving features in well-constrained regions. These results indicate that the diffusion prior complements the physical gradient by restoring structures in regions where the seismic sensitivity is weak.

\begin{figure}[H]
\centering
\includegraphics[width=\columnwidth]{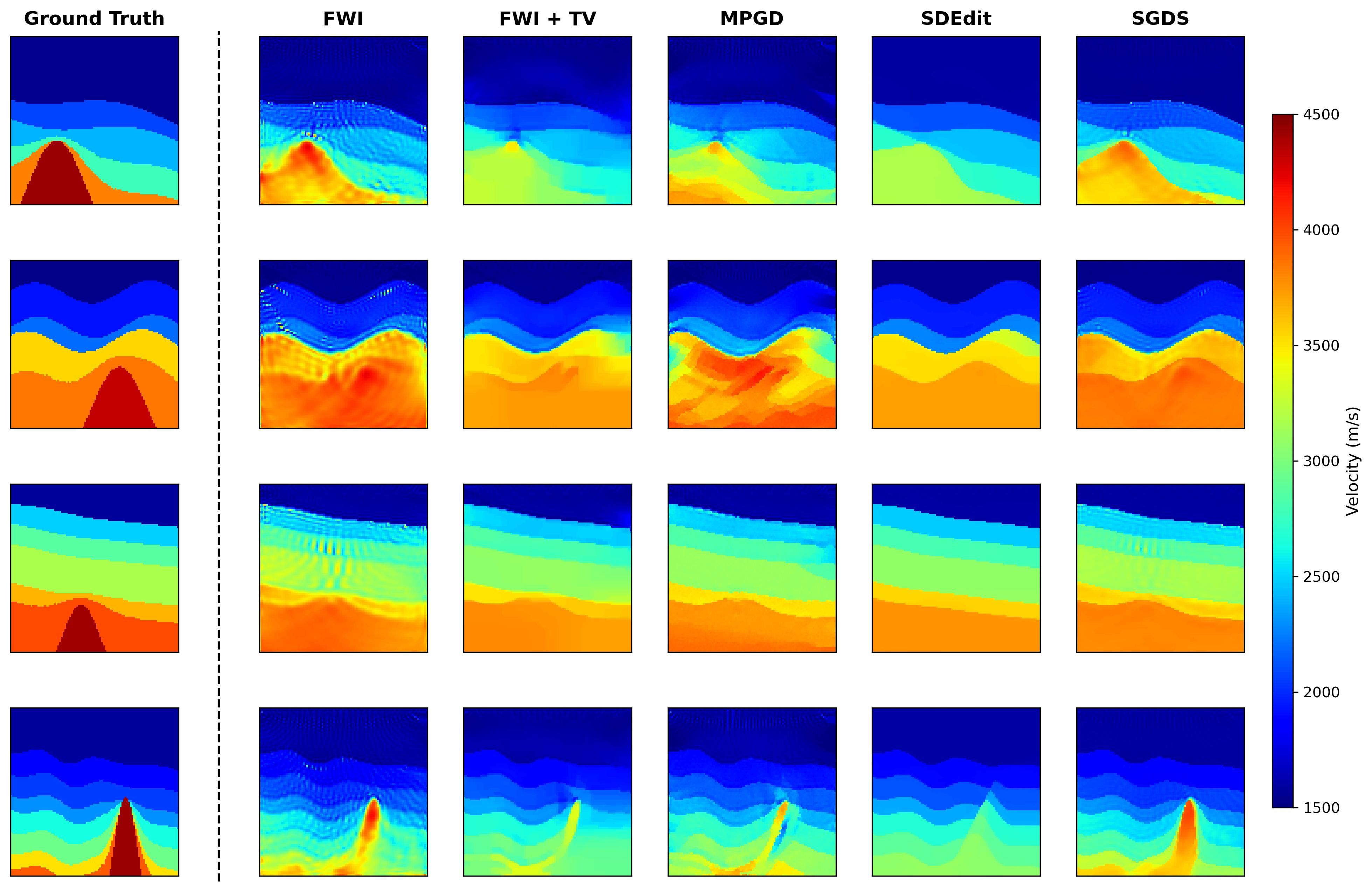}
\caption{Salt body inversion results for four synthetic models. Columns show the ground truth, conventional FWI using an $\ell_2$ misfit, FWI with total variation (TV) regularization, Manifold-Preserving Guided Diffusion (MPGD), SDEdit, and Split Gibbs Diffusion Sampling (SGDS). Conventional FWI produces strong artifacts and fails to reconstruct the salt geometry due to cycle skipping and limited sensitivity within the high-velocity salt region. TV regularization improves stability but oversmooths structural contrasts and reduces the salt extent. MPGD and SDEdit partially improve reconstruction but remain affected by optimization bias or insufficient physical guidance. In contrast, SGDS better recovers the salt geometry and velocity, preserving sharper boundaries and producing models that more closely follow the ground truth. These results suggest that diffusion-based regularization restores structural information that is poorly constrained by the seismic gradient.
}
\label{fig:grid_Salt}
\end{figure}

\begin{figure}[H]
\centering
\includegraphics[width=\columnwidth]{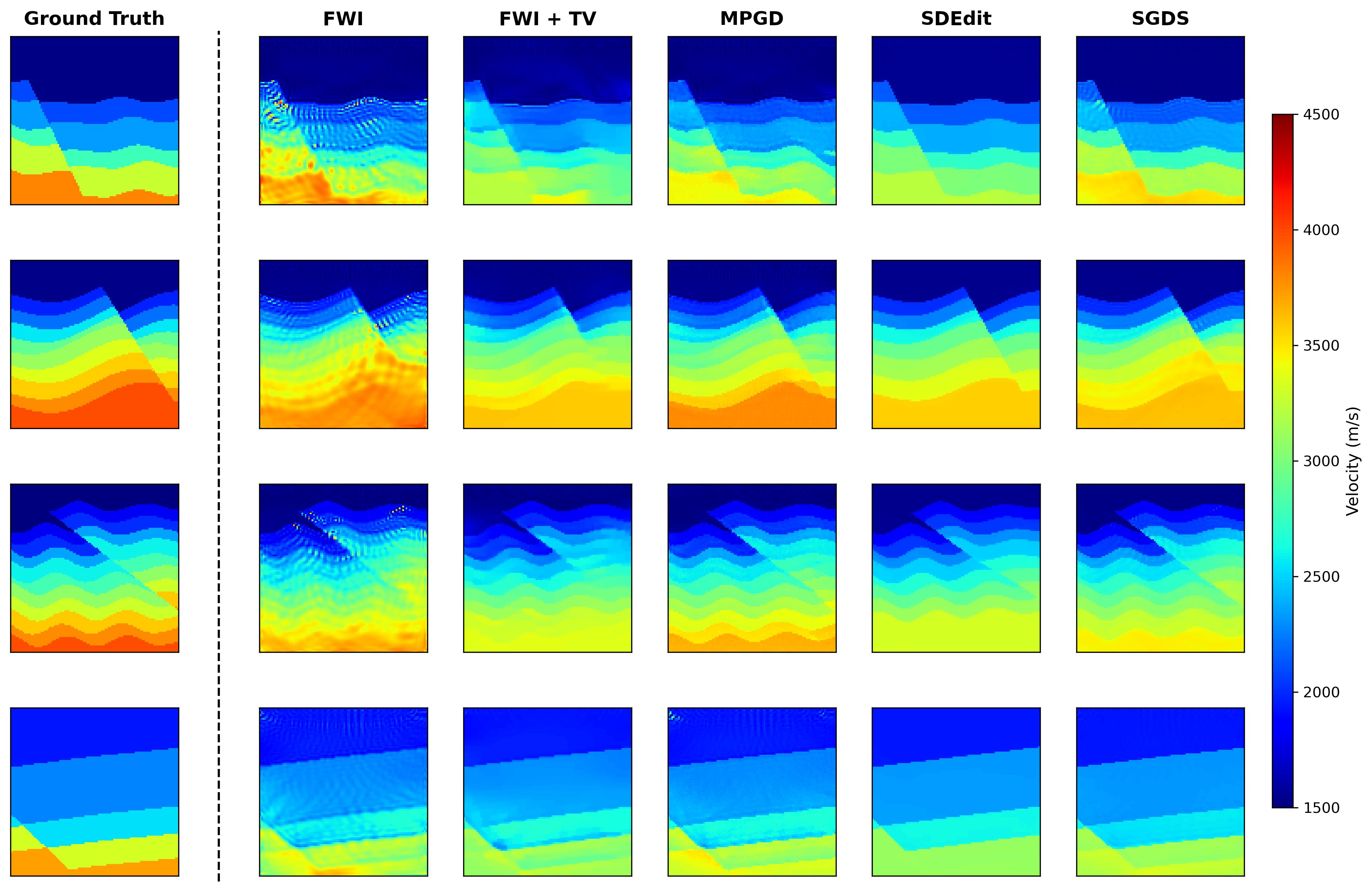}
\caption{Faulted velocity model inversion results for four synthetic examples. Conventional FWI struggles to reconstruct the fault geometry and produces strong artifacts due to cycle skipping and nonlinear inversion effects. TV regularization reduces instability but oversmooths the fault interface and suppresses velocity contrasts. MPGD partially reconstructs the fault structure but exhibits distortions in deeper regions. SDEdit produces smoother models but lacks sufficient structural accuracy. In contrast, SGDS better recovers both the fault geometry and velocity distribution, preserving sharper discontinuities and producing models more consistent with the ground truth. These results indicate that diffusion-guided inversion improves recovery of discontinuous geological structures.}
\label{fig:grid_Fault}
\end{figure}

\begin{figure}[H]
\centering
\includegraphics[width=\columnwidth]{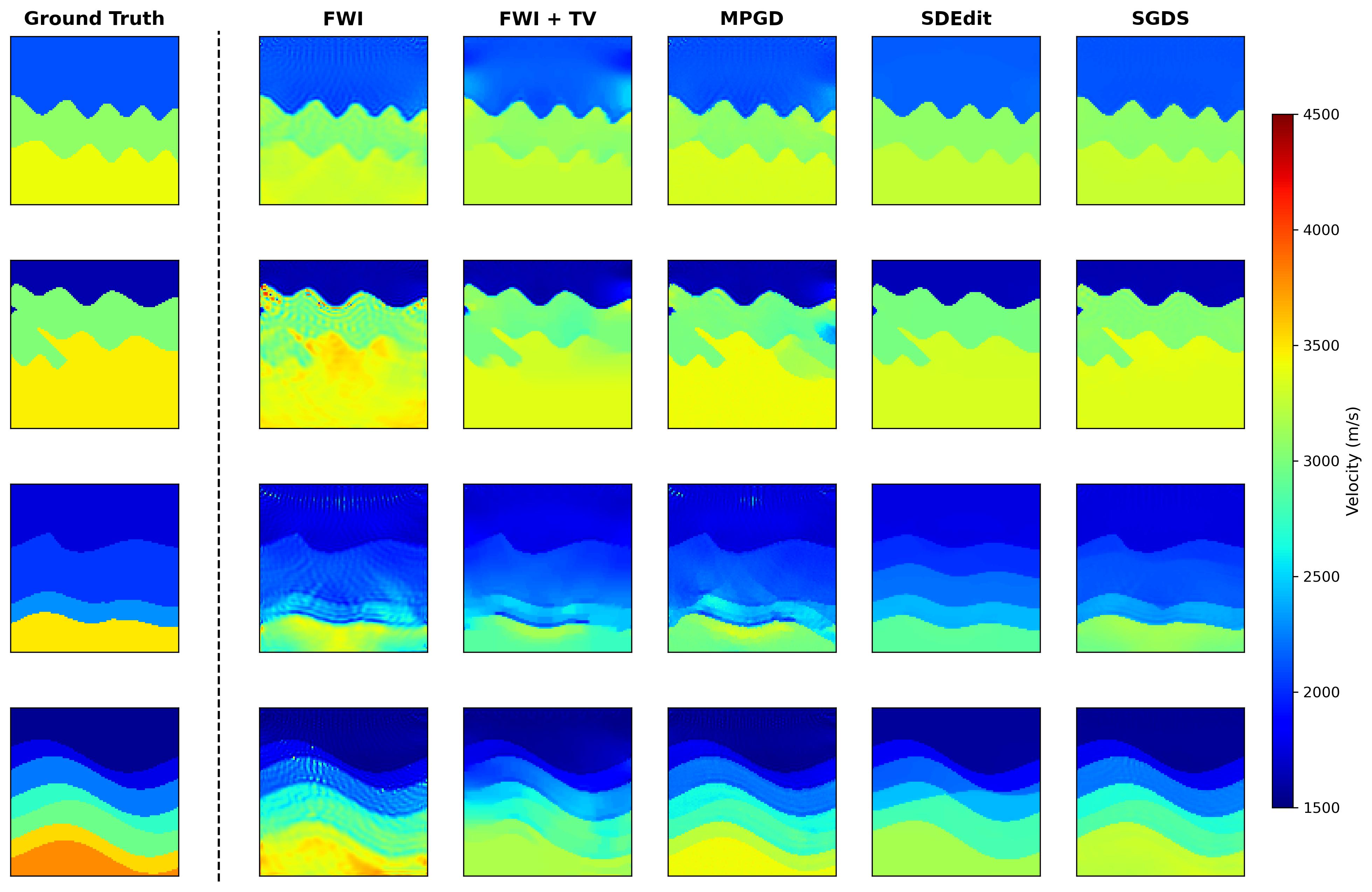}
\caption{Layered velocity model inversion results for four synthetic examples. Conventional FWI produces oscillatory artifacts and inaccurate layer boundaries, particularly in deeper regions where seismic sensitivity is reduced. TV regularization stabilizes the inversion but suppresses velocity contrasts and smooths structural interfaces. MPGD improves structural continuity but introduces localized distortions due to the interaction between optimization and generative constraints. SDEdit produces smooth models but lacks sufficient data consistency to fully recover layer geometry. In contrast, SGDS better recovers layered structures and preserves velocity contrasts across depth. These results indicate that the diffusion prior can restore high-wavenumber structural information while maintaining consistency with seismic observations.}
\label{fig:grid_Layer}
\end{figure}

Across all three scenarios (Figures~\ref{fig:grid_Salt}, \ref{fig:grid_Fault}, and \ref{fig:grid_Layer}), SGDS gives the most consistent overall behavior among the tested couplings, while SDEdit performs well in moderately complex regions. MPGD preserves manifold-conforming structures in some layered cases but can introduce localized distortions. These findings suggest that generative priors, when guided appropriately, can serve as an adaptive and interpretable form of regularization alongside classical techniques.

Next, we benchmark the effectiveness of the Split Gibbs Diffusion Sampling (SGDS) method as a learned regularization prior for full waveform inversion (FWI), comparing it with classical approaches such as total variation (TV) regularization and conventional $L_2$-based inversion. Unlike previous experiments involving multiple guidance strategies, here we select SGDS because it gives the most consistent performance in the controlled GeoFWI tests and has a clearer alternating likelihood-prior structure. A full benchmark-scale comparison of MPGD, SDEdit, and SGDS would require separate tuning of each method and is left for future work.

The diffusion prior was trained solely on the GeoFWI dataset using $100 \times 100$ velocity patches. To adapt this model to benchmark datasets of different spatial resolutions, we apply a patchwise reconstruction scheme with 10\% overlap, allowing us to generate high-resolution predictions while preserving local consistency. Because Marmousi and Overthrust are not used in diffusion training, these experiments test benchmark-scale transfer of the GeoFWI-trained prior to larger and structurally distinct velocity models.

\subsection{Effect of diffusion initialization level in SGDS}
\begin{figure}[H]
\centering
\includegraphics[width=\columnwidth]{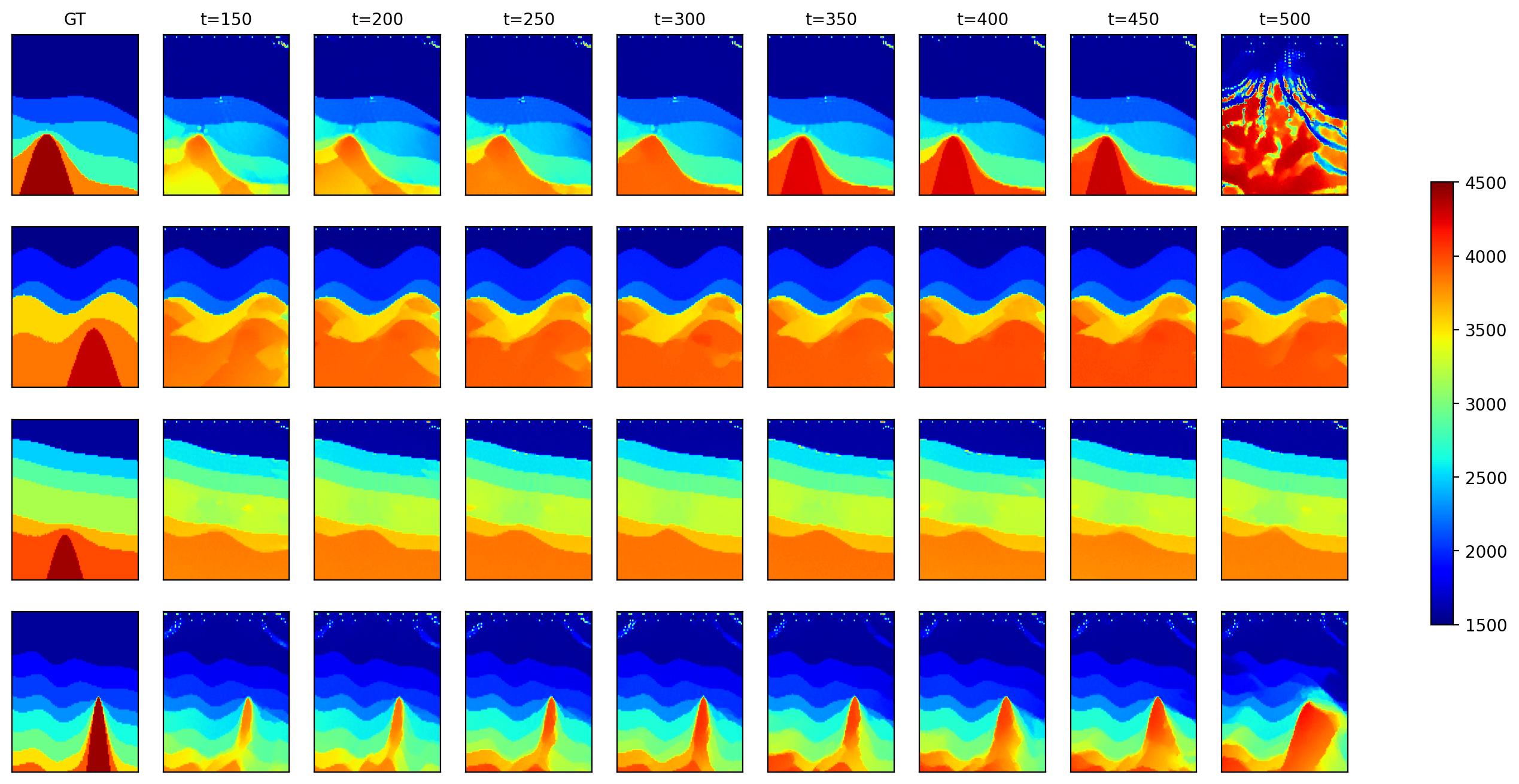}
\caption{Effect of diffusion initialization level in the SGDS algorithm for four salt body models. The first column shows the ground truth, and subsequent columns show inversion results using SGDS initialized at different diffusion times $t$. Smaller values of $t$ introduce limited perturbations and preserve the current FWI estimate, resulting in incomplete recovery of salt geometry. Intermediate values ($t \approx 300$--$400$) provide sufficient exploration of the learned generative manifold while maintaining consistency with seismic constraints, producing the strongest reconstructions among the tested settings. Excessively large values (e.g., $t=500$) introduce strong noise that degrades reconstruction quality by destroying physically meaningful structures. These results indicate that the diffusion initialization level controls the balance between escaping local minima and preserving data consistency, with intermediate noise levels performing best in these representative cases.
}
\label{fig:nosie-sgds}
\end{figure}

We investigated the effect of the diffusion initialization level on inversion performance in the SGDS framework. Figure~\ref{fig:nosie-sgds} and Table~\ref{tab:quantitative} show that the diffusion start time controls the trade-off between escaping local minima and preserving physically meaningful structure. Initializing at intermediate noise levels improves reconstruction quality by enabling exploration while retaining information from the current FWI estimate. In our representative GeoFWI cases, SGDS ($t{=}350$) yields the strongest average performance among the tested noise levels, achieving the highest PSNR and the lowest relative $\ell_2$ error with moderate computational cost. In contrast, overly large noise levels degrade accuracy due to loss of physical information, while insufficient noise limits the ability of the diffusion prior to correct inversion errors.

\begin{table}[t]
\centering
\caption{Quantitative comparison on velocity model recovery for 4 GeoFWI test cases per category.
\textbf{Bold}: best; \underline{Underline}: second best.}
\label{tab:quantitative}
\resizebox{\textwidth}{!}{%
\begin{tabular}{l ccc ccc ccc}
\toprule
 & \multicolumn{3}{c}{Layer} & \multicolumn{3}{c}{Salt} & \multicolumn{3}{c}{Fault} \\
\cmidrule(lr){2-4} \cmidrule(lr){5-7} \cmidrule(lr){8-10}
Method & PSNR & SSIM & Rel.~$\ell_2$ & PSNR & SSIM & Rel.~$\ell_2$ & PSNR & SSIM & Rel.~$\ell_2$ \\
\midrule
FWI       & 25.75 & 0.7982 & 0.0630 & 22.38 & 0.7480 & 0.0872 & 25.08 & 0.7749 & 0.0626 \\
FWI + TV  & 23.22 & 0.8523 & 0.0857 & 19.83 & 0.8279 & 0.1214 & 22.10 & 0.8314 & 0.0879 \\
MPGD      & 25.80 & 0.8738 & 0.0652 & 20.01 & 0.7381 & 0.1142 & 25.79 & 0.8762 & 0.0578 \\
SDEdit    & 23.47 & 0.8723 & 0.0835 & 19.76 & 0.8333 & 0.1218 & 22.08 & 0.8356 & 0.0880 \\
\midrule
SGDS ($\sigma_1{=}150$) & 24.90 & 0.8441 & 0.0695 & 21.26 & 0.7898 & 0.0987 & 25.06 & 0.8385 & 0.0641 \\
SGDS ($\sigma_1{=}200$) & 25.56 & \underline{0.8475} & 0.0638 & 22.07 & 0.8041 & 0.0894 & 25.95 & \underline{0.8412} & 0.0585 \\
SGDS ($\sigma_1{=}250$) & 25.19 & 0.8069 & 0.0715 & 22.60 & 0.8081 & 0.0829 & 26.05 & 0.8282 & 0.0582 \\
SGDS ($\sigma_1{=}300$) & \underline{25.90} & 0.8315 & \underline{0.0640} & 23.28 & \textbf{0.8201} & 0.0755 & \underline{26.04} & \textbf{0.9085} & \underline{0.0567} \\
SGDS ($\sigma_1{=}350$) & \textbf{28.62} & \textbf{0.9512} & \textbf{0.0459} & \textbf{23.60} & 0.8109 & \textbf{0.0726} & \textbf{27.57} & 0.8332 & \textbf{0.0484} \\
SGDS ($\sigma_1{=}400$) & 25.01 & 0.8025 & 0.0748 & \underline{23.50} & 0.8095 & \underline{0.0730} & 25.20 & 0.8153 & 0.0646 \\
SGDS ($\sigma_1{=}450$) & 24.28 & 0.7783 & 0.0859 & 23.41 & \underline{0.8140} & 0.0741 & 25.67 & 0.8234 & 0.0605 \\
SGDS ($\sigma_1{=}500$) & 23.96 & 0.7761 & 0.0950 & 18.75 & 0.6502 & 0.1677 & 25.42 & 0.8231 & 0.0636 \\
\bottomrule
\end{tabular}%
}
\end{table}

The expanded 18-case statistics in Table~\ref{tab:geofwi_heldout_stats} provide a more conservative summary than the representative sweep above. SGDS gives the best overall mean PSNR, SSIM, and relative $\ell_2$ error across all complete cases. At the category level, SGDS is strongest on the Salt and Fault subsets for all three metrics. On the Layer subset, SGDS gives the best mean PSNR and relative $\ell_2$ error, while SDEdit gives a slightly higher mean SSIM (0.929 versus 0.926). This result supports the use of SGDS as the most consistent of the tested couplings, while also showing that individual metrics can favor different diffusion-guidance strategies.

\begin{table}[H]
\centering
\caption{Expanded held-out GeoFWI statistics over 18 complete inversions (6 examples per category). Values are mean $\pm$ standard deviation. Higher PSNR and SSIM are better; lower relative $\ell_2$ error is better.}
\label{tab:geofwi_heldout_stats}
\begin{tabular}{lccc}
\toprule
\textbf{Method} & \textbf{PSNR} & \textbf{SSIM} & \textbf{Rel.~$\ell_2$} \\
\midrule
FWI $L_2$ & $19.96 \pm 3.91$ & $0.658 \pm 0.134$ & $0.073 \pm 0.038$ \\
FWI $L_2$+TV & $19.43 \pm 3.35$ & $0.798 \pm 0.112$ & $0.081 \pm 0.043$ \\
MPGD & $19.99 \pm 4.03$ & $0.744 \pm 0.141$ & $0.080 \pm 0.051$ \\
SDEdit & $19.76 \pm 2.99$ & $0.868 \pm 0.068$ & $0.078 \pm 0.040$ \\
SGDS & $\mathbf{21.56 \pm 3.51}$ & $\mathbf{0.884 \pm 0.045}$ & $\mathbf{0.063 \pm 0.032}$ \\
\bottomrule
\end{tabular}
\end{table}

\subsection{Marmousi benchmark}
The Marmousi model represents a challenging inversion scenario due to its strong lateral velocity variations, dipping sedimentary layers, and complex structural geometry. The diffusion prior used in this work was trained on $100\times100$ velocity patches from the GeoFWI dataset and applied to the Marmousi model ($321\times901$ grid points) using a patchwise reconstruction strategy with overlapping windows. All inversion methods were initialized from the same smooth velocity model and employed a multiscale frequency continuation schedule of 10, 15, 20, and 25 Hz.

\begin{figure}[H]
\centering
\includegraphics[width=0.85\columnwidth]{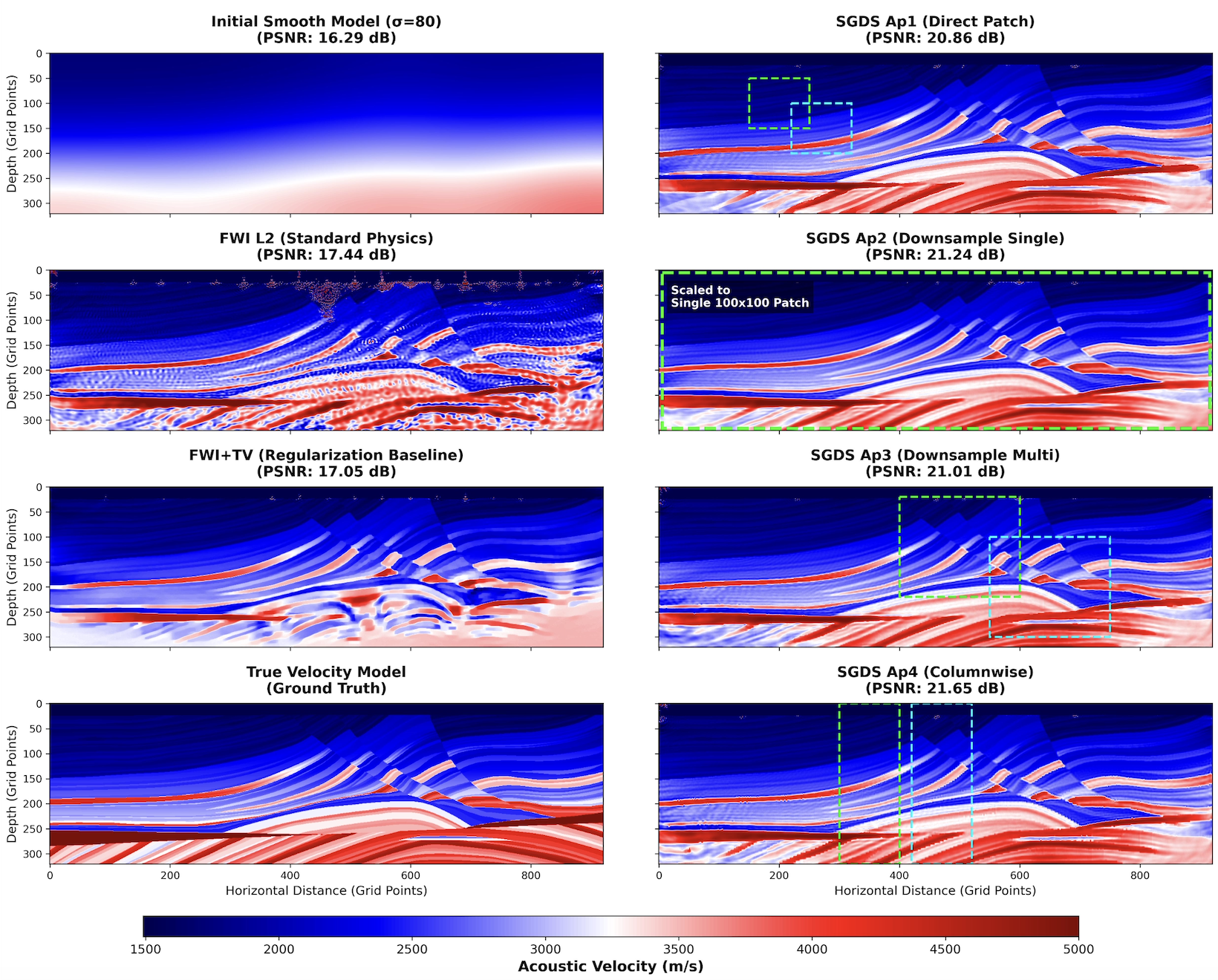}
\caption{Marmousi inversion results. (a) Initial smooth velocity model used for inversion. (b) Conventional FWI using $\ell_2$ misfit, which exhibits cycle-skipping artifacts and inaccurate velocity recovery, particularly at depth. (c) FWI with total variation regularization, which improves stability but suppresses velocity contrasts and oversmooths geological interfaces. (d) Diffusion-guided inversion using SGDS, which better recovers layered structures and preserves velocity contrasts. (e) Ground truth velocity model. Vertical black lines indicate locations of velocity profiles shown in the right column. The velocity profiles show that the diffusion-guided inversion more closely follows the ground truth across most depths, particularly in poorly illuminated regions where conventional inversion exhibits large errors.}
\label{fig:marmousi}
\end{figure}

Figure~\ref{fig:marmousi} compares the reconstructed velocity models obtained using conventional $\ell_2$ full waveform inversion (FWI), total variation (TV) regularized inversion, and diffusion-guided inversion using the Split Gibbs Diffusion Sampling (SGDS) framework. The conventional $\ell_2$ inversion suffers from cycle skipping and produces significant velocity errors, particularly in deeper regions where seismic illumination is limited. TV regularization improves stability by suppressing oscillatory artifacts but introduces excessive smoothing, reducing velocity contrast and blurring important geological interfaces.

In contrast, the diffusion-guided inversion produces a velocity model that more closely follows the ground truth, preserving continuous stratigraphic layering and recovering sharper velocity contrasts. The learned diffusion prior encourages the inversion to remain near a manifold of geologically plausible models, reducing the tendency to converge toward physically inconsistent solutions.

To quantitatively evaluate reconstruction accuracy, vertical velocity profiles at representative lateral locations are shown in the right column of Figure~\ref{fig:marmousi}. The conventional $\ell_2$ inversion exhibits large deviations from the ground truth, especially below approximately 600~m depth, where the seismic gradient becomes weak. TV regularization improves stability but underestimates velocity contrasts due to its smoothing effect. In contrast, the diffusion-guided inversion more closely follows the ground-truth velocity profile across most depths, better recovering both the long-wavelength background and high-wavenumber structural variations. This result suggests that the diffusion prior complements the physical gradient by restoring structural components that are poorly constrained by the seismic data alone.

\begin{figure}[H]
\centering
\includegraphics[width=0.72\columnwidth]{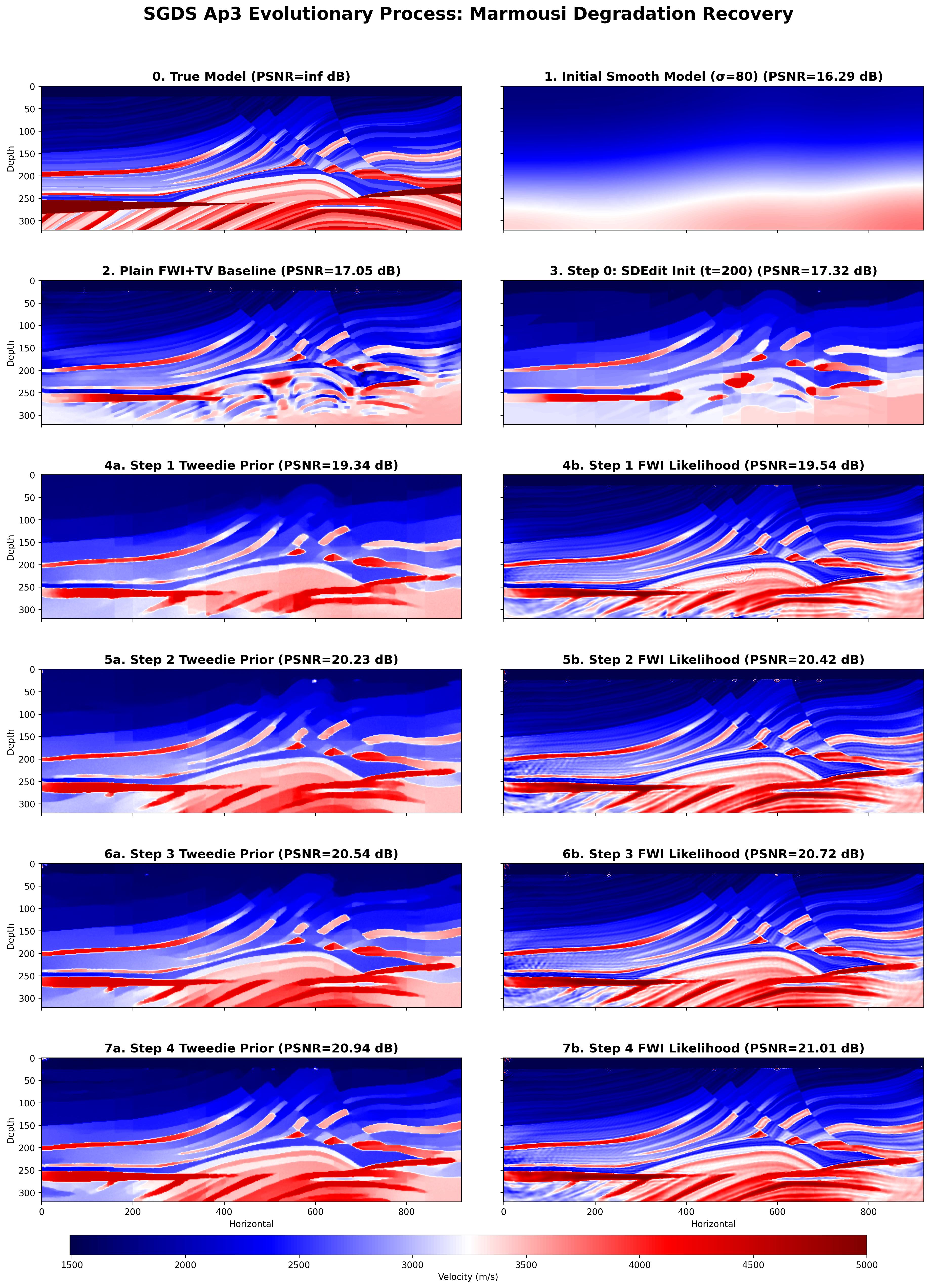}
\caption{Step-by-step Marmousi SGDS evolution using the downsample-multi deployment mode. The sequence shows how alternating diffusion-prior denoising and deterministic FWI likelihood updates progressively remove implausible oscillations while preserving data-consistent structure. This figure is included to make the SGDS mechanism visible rather than only reporting the final model.}
\label{fig:marmousi_ap3_evolution}
\end{figure}

\begin{figure}[H]
\centering
\includegraphics[width=\columnwidth]{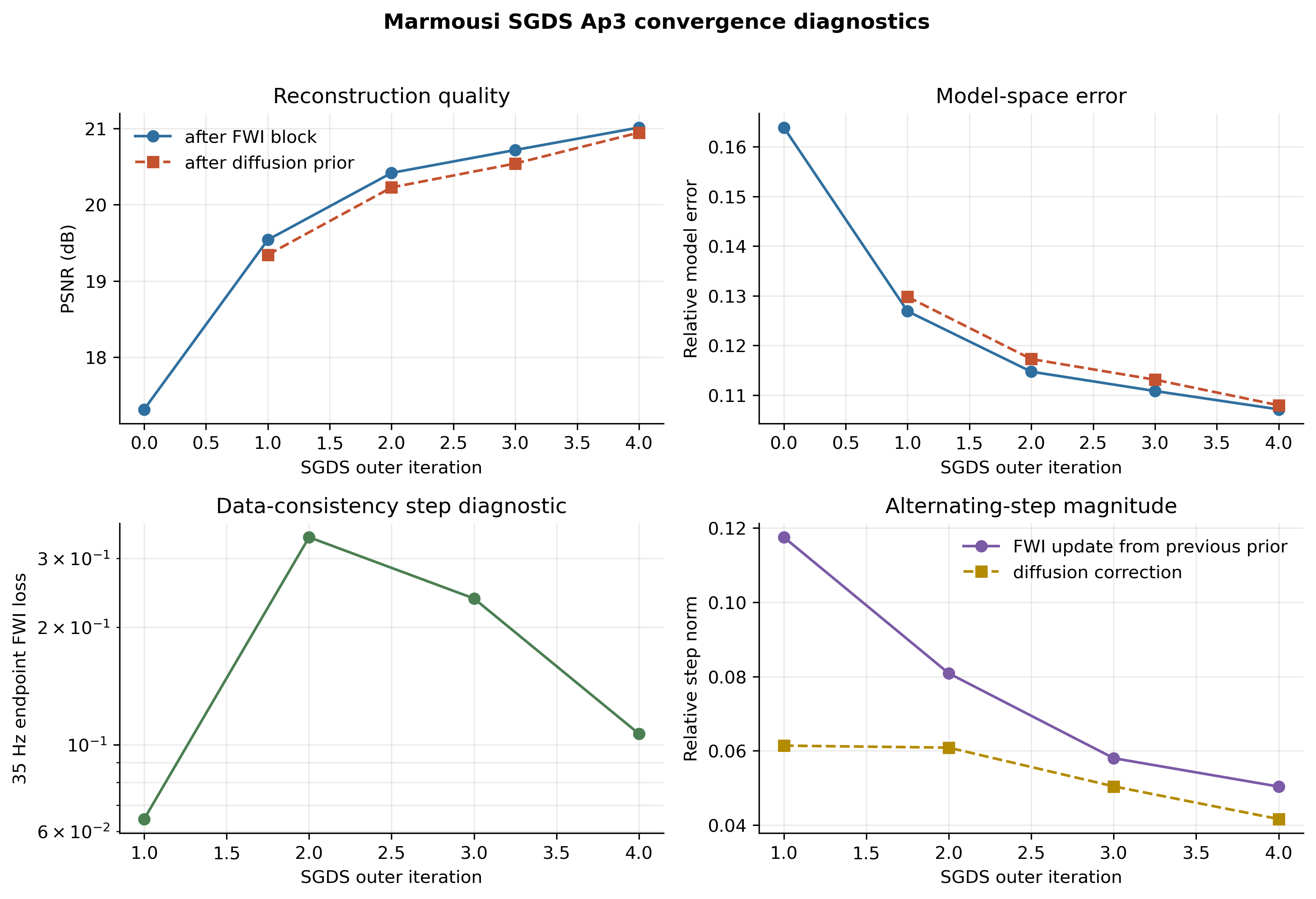}
\caption{Marmousi SGDS Ap3 convergence diagnostics computed from saved outer-loop states. The FWI blocks progressively improve PSNR and reduce relative $\ell_2$ model error, while the diffusion-prior correction norm decreases with iteration. The 35 Hz endpoint loss is shown as a data-consistency diagnostic; it is not expected to be globally monotone across outer loops because the prior step intentionally modifies the model before the next FWI refinement.}
\label{fig:marmousi_sgds_convergence}
\end{figure}

To further assess the stability of the alternating workflow, Figure~\ref{fig:marmousi_sgds_convergence} reports convergence diagnostics for the same Marmousi SGDS Ap3 run. The PSNR increases from 17.32 dB after the SDEdit initialization to 21.01 dB after the fourth FWI block, while the relative $\ell_2$ model error decreases from 0.164 to 0.107. The diffusion-prior correction norm also decreases across outer iterations, indicating that the denoising step becomes less aggressive as the iterate approaches a more stable structural regime. The high-frequency FWI loss is not monotone across outer loops because each diffusion-prior step changes the model before the next likelihood update. Thus, these diagnostics support a local alternating-stability interpretation rather than a global monotone-convergence claim.

\subsection{Overthrust benchmark}

The Overthrust model provides a particularly challenging inversion scenario due to its strong velocity contrasts, deep high-velocity layers, and complex structural geometry. The diffusion prior was trained only on GeoFWI patches and applied to the Overthrust model using a patchwise reconstruction strategy, making this experiment a benchmark-scale transfer test on a larger and structurally distinct model.

\begin{figure}[H]
\centering
\includegraphics[width=0.85\columnwidth]{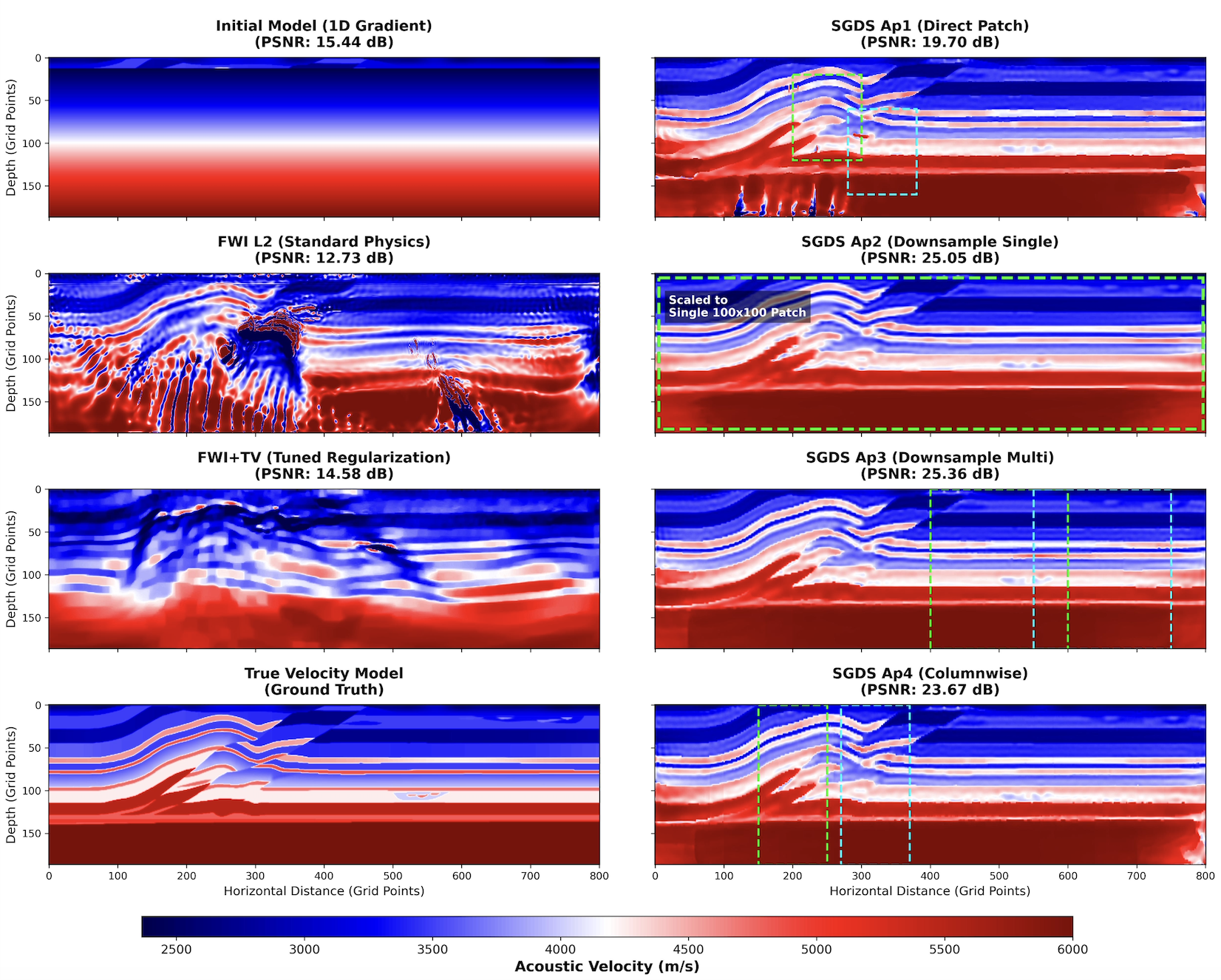}
\caption{Overthrust inversion results. (a) Initial smooth velocity model. (b) Conventional $\ell_2$ inversion showing large velocity errors and cycle skipping at depth. (c) TV-regularized inversion, which improves stability but oversmooths velocity contrasts. (d) Diffusion-guided inversion using SGDS, which better recovers layered structures and deep velocity contrasts. (e) Ground truth model. Vertical black lines indicate locations of velocity profiles shown in the right column. The velocity profiles show that diffusion-guided inversion more closely follows the ground truth, particularly in deep regions where conventional inversion exhibits large errors.
}
\label{fig:overthrust}
\end{figure}

Figure~\ref{fig:overthrust} compares inversion results obtained using conventional $\ell_2$ full waveform inversion, total variation (TV) regularized inversion, and diffusion-guided inversion using the SGDS framework. Conventional $\ell_2$ inversion produces significant velocity errors, particularly in deeper regions below approximately 2000~m depth, where seismic illumination is limited. These errors arise from cycle skipping and insufficient sensitivity of the seismic gradient to deep structures.

TV regularization improves inversion stability and suppresses oscillatory artifacts but introduces excessive smoothing, which reduces velocity contrasts and suppresses important structural features. In contrast, the diffusion-guided inversion better recovers layered structures and preserves sharper velocity contrasts throughout the model.

Vertical velocity profiles at representative lateral positions are shown in the right column of Figure~\ref{fig:overthrust}. Conventional $\ell_2$ inversion significantly overestimates velocity in deep regions and fails to recover correct layer boundaries. TV regularization reduces instability but introduces systematic bias due to smoothing. In contrast, the diffusion-guided inversion more closely follows the ground-truth velocity profile across most depths, better recovering both the deep high-velocity region and intermediate structural variations. This result suggests that the diffusion prior restores structural components that are poorly constrained by the seismic data alone and improves robustness against cycle skipping.

Because the Overthrust experiment uses a deliberately simple linear smooth initial model, it also provides a useful test of whether the SGDS improvement is mainly inherited from the deterministic seed or from the diffusion initialization. We therefore performed two additional Ap3/downsample-multi ablations while keeping the SGDS schedule, FWI frequency continuation, patch deployment, and diffusion model fixed. First, we removed the initial SDEdit step and started the alternating FWI--Tweedie loop directly from the FWI+TV seed. Second, we kept the SDEdit initialization but replaced the FWI+TV seed by the weaker $L_2$-FWI seed. These ablations are reported before the final benchmark polishing step, so the absolute SGDS number differs from the final Overthrust result in Table~\ref{tab:benchmark_metrics}.

Table~\ref{tab:overthrust_ablation} shows that the initial SDEdit step is important in this poor-starting-model case. Without SDEdit initialization, the SGDS loop improves structural similarity visually but gives only limited RMSE/NRMSE improvement over the smooth starting model. Starting SGDS from the weaker $L_2$-FWI seed still improves the $L_2$-FWI result, indicating that the method is not simply reproducing TV regularization. However, the best result is obtained when a more stable FWI+TV seed is combined with SDEdit initialization. This supports the interpretation that SGDS remains basin-dependent in nonlinear FWI: the deterministic FWI seed, SDEdit basin-entry step, and alternating FWI--Tweedie loop play complementary roles.

\begin{table}[t]
\centering
\caption{Overthrust SGDS ablation under the Ap3/downsample-multi setting. The ablations use the same linear smooth initial model, SGDS schedule, patch deployment, and diffusion model. RMSE is reported in m/s.}
\label{tab:overthrust_ablation}
\begin{tabular}{lrrrr}
\toprule
\textbf{Method} & \textbf{PSNR} & \textbf{RMSE} & \textbf{NRMSE} & \textbf{Rel. $L_2$} \\
\midrule
Smooth linear initial model & 15.44 & 614.5 & 0.1691 & 0.1323 \\
FWI $L_2$ seed & 12.73 & 839.2 & 0.2309 & 0.1806 \\
FWI $L_2$+TV seed & 14.58 & 678.1 & 0.1865 & 0.1459 \\
SGDS from FWI $L_2$ seed & 16.25 & 559.7 & 0.1540 & 0.1204 \\
SGDS from FWI+TV seed, no SDEdit init & 15.40 & 617.0 & 0.1698 & 0.1328 \\
SGDS from FWI+TV seed with SDEdit init & \textbf{23.83} & \textbf{233.9} & \textbf{0.0644} & \textbf{0.0503} \\
\bottomrule
\end{tabular}
\end{table}

\begin{table}[t]
\centering
\caption{Benchmark-scale quantitative comparison. Higher PSNR and SSIM are better; lower RMSE and NRMSE are better. RMSE is reported in m/s.}
\label{tab:benchmark_metrics}
\begin{tabular}{llrrrr}
\toprule
\textbf{Model} & \textbf{Method} & \textbf{PSNR} & \textbf{SSIM} & \textbf{RMSE} & \textbf{NRMSE} \\
\midrule
Marmousi & Initial & 16.29 & 0.6198 & 537.7 & 0.1532 \\
Marmousi & FWI $L_2$ & 17.44 & 0.5947 & 471.2 & 0.1343 \\
Marmousi & FWI $L_2$+TV & 17.05 & 0.6851 & 493.1 & 0.1405 \\
Marmousi & SGDS & 21.01 & 0.7867 & 312.4 & 0.0890 \\
\midrule
Overthrust & Initial & 15.44 & 0.5414 & 614.5 & 0.1691 \\
Overthrust & FWI $L_2$ & 12.73 & 0.4309 & 839.2 & 0.2309 \\
Overthrust & FWI $L_2$+TV & 14.58 & 0.4849 & 678.1 & 0.1865 \\
Overthrust & SGDS & 25.36 & 0.8135 & 196.1 & 0.0540 \\
\bottomrule
\end{tabular}
\end{table}

\subsection{Noise robustness on Marmousi}
To test whether the learned prior improves robustness beyond clean synthetic data, we repeated the Marmousi experiment after adding Gaussian noise to the observed data. Table~\ref{tab:noise_degradation} reports PSNR degradation for clean data and for 20, 10, and 5 dB SNR. SGDS remains the best method for clean, 20 dB, and 10 dB data. At the extreme 5 dB stress-test level, all methods are substantially degraded, and the likelihood update becomes too noisy to reliably guide the alternating scheme. Thus, SGDS improves moderate-noise robustness, while the 5 dB case identifies a regime where noise-aware schedules or likelihood models become important.

\begin{table}[t]
\centering
\caption{Marmousi PSNR degradation under additive Gaussian noise.}
\label{tab:noise_degradation}
\begin{tabular}{lrrrr}
\toprule
\textbf{Method} & \textbf{Clean} & \textbf{20 dB} & \textbf{10 dB} & \textbf{5 dB} \\
\midrule
FWI $L_2$ & 17.44 & 14.73 & 8.63 & 4.36 \\
FWI $L_2$+TV & 17.05 & 15.60 & 11.46 & 6.42 \\
SGDS & 21.01 & 19.70 & 13.60 & 2.88 \\
\bottomrule
\end{tabular}
\end{table}

\begin{figure}[H]
\centering
\includegraphics[width=\columnwidth]{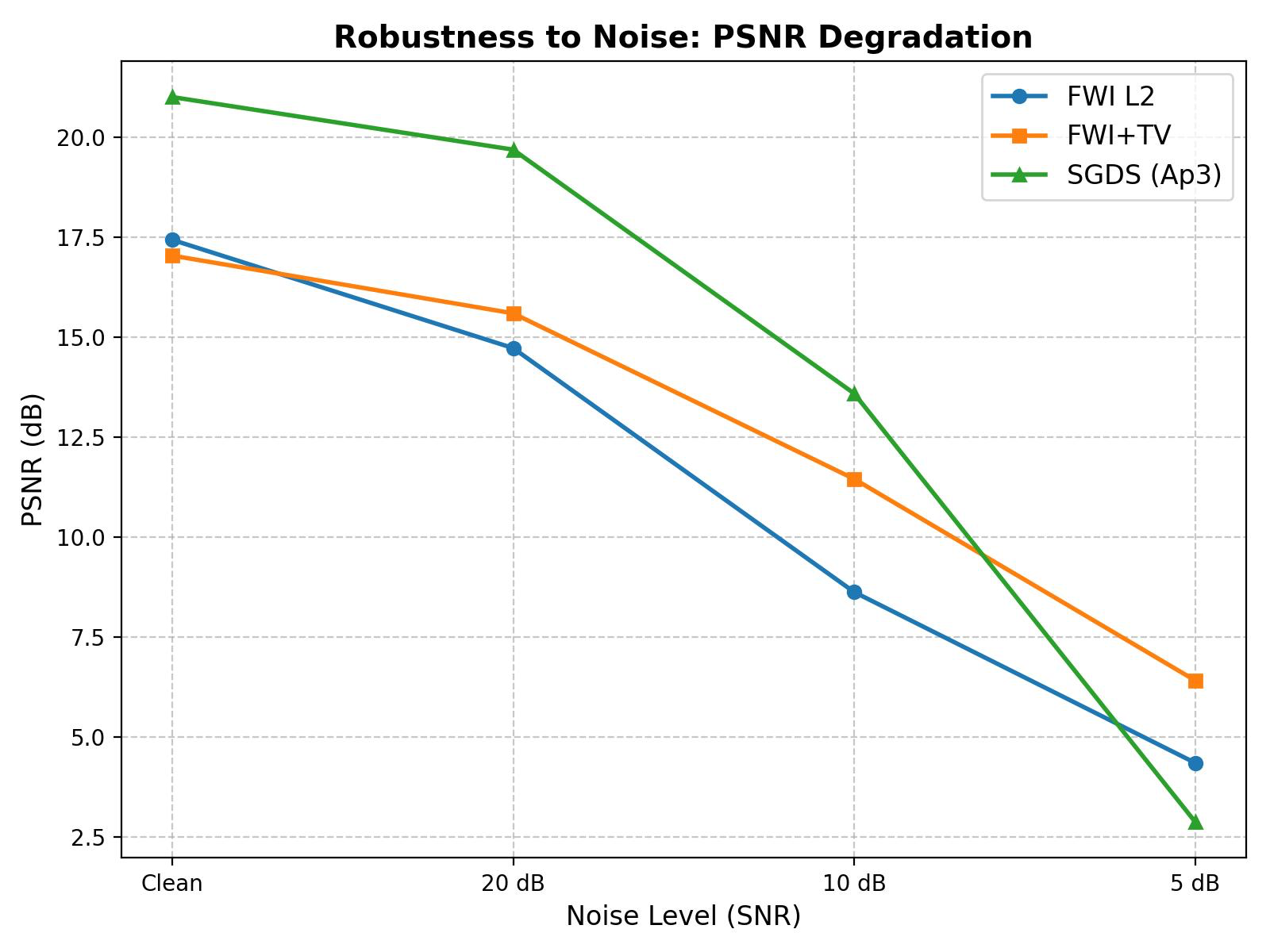}
\caption{Noise-degradation curve for the Marmousi experiment. SGDS gives higher PSNR than the classical baselines for clean, 20 dB, and 10 dB observations. The 5 dB stress test shows the robustness boundary where the likelihood update becomes too noise dominated for the current schedule.}
\label{fig:marmousi_noise_degradation}
\end{figure}

\subsection{Sigsbee2A salt stress test}
The Sigsbee2A model is included as a harder salt-dominated benchmark with a poor smooth starting model, strong velocity contrast, and severe subsalt illumination deficit. This experiment addresses a different question from the GeoFWI examples: whether a GeoFWI-trained prior can still be useful when the target contains a large, coherent salt body whose geometry is not well recovered by direct gradient-based inversion. In this setting, SGDS is used as a structural proposal mechanism rather than as a single black-box denoiser. The prior output is converted into a salt mask, the salt region is flooded with a physically plausible salt velocity, and deterministic FWI refines the surrounding background. Therefore, this experiment should be interpreted as a workflow-level stress test rather than a direct denoiser-only comparison.

\begin{figure}[H]
\centering
\includegraphics[width=0.82\columnwidth]{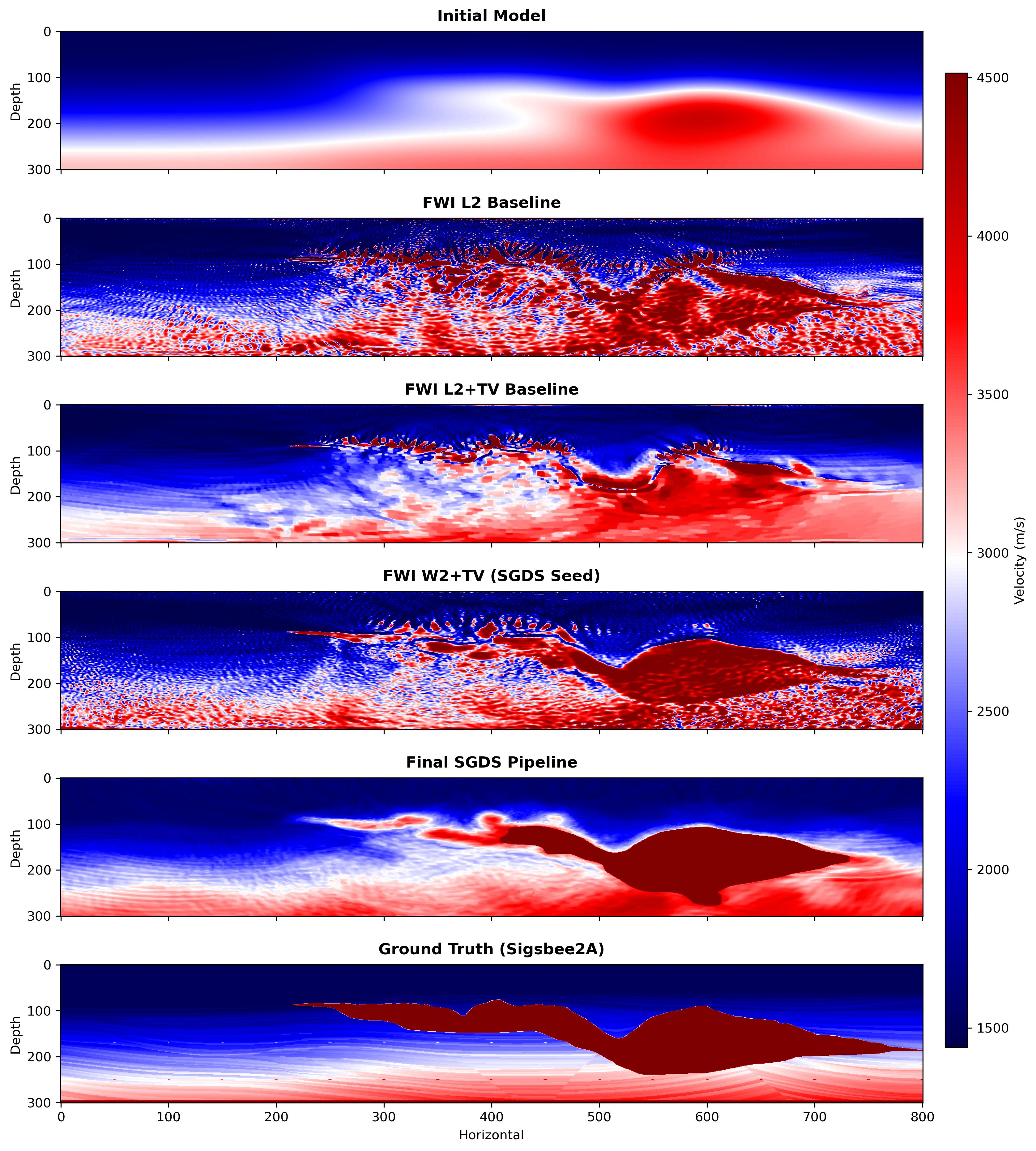}
\caption{Sigsbee2A salt benchmark. Rows show the smooth initial model, conventional $L_2$ FWI, $L_2$+TV FWI, a transport-informed seed used to initialize the SGDS workflow, the final SGDS-assisted reconstruction, and the ground-truth model. The example is included as a poor-starting-model workflow stress test: direct FWI baselines fail to recover a coherent salt body, whereas the SGDS-assisted workflow recovers the dominant salt geometry and a cleaner background trend.}
\label{fig:sigsbee2a_final}
\end{figure}

\begin{figure}[H]
\centering
\includegraphics[width=0.78\columnwidth]{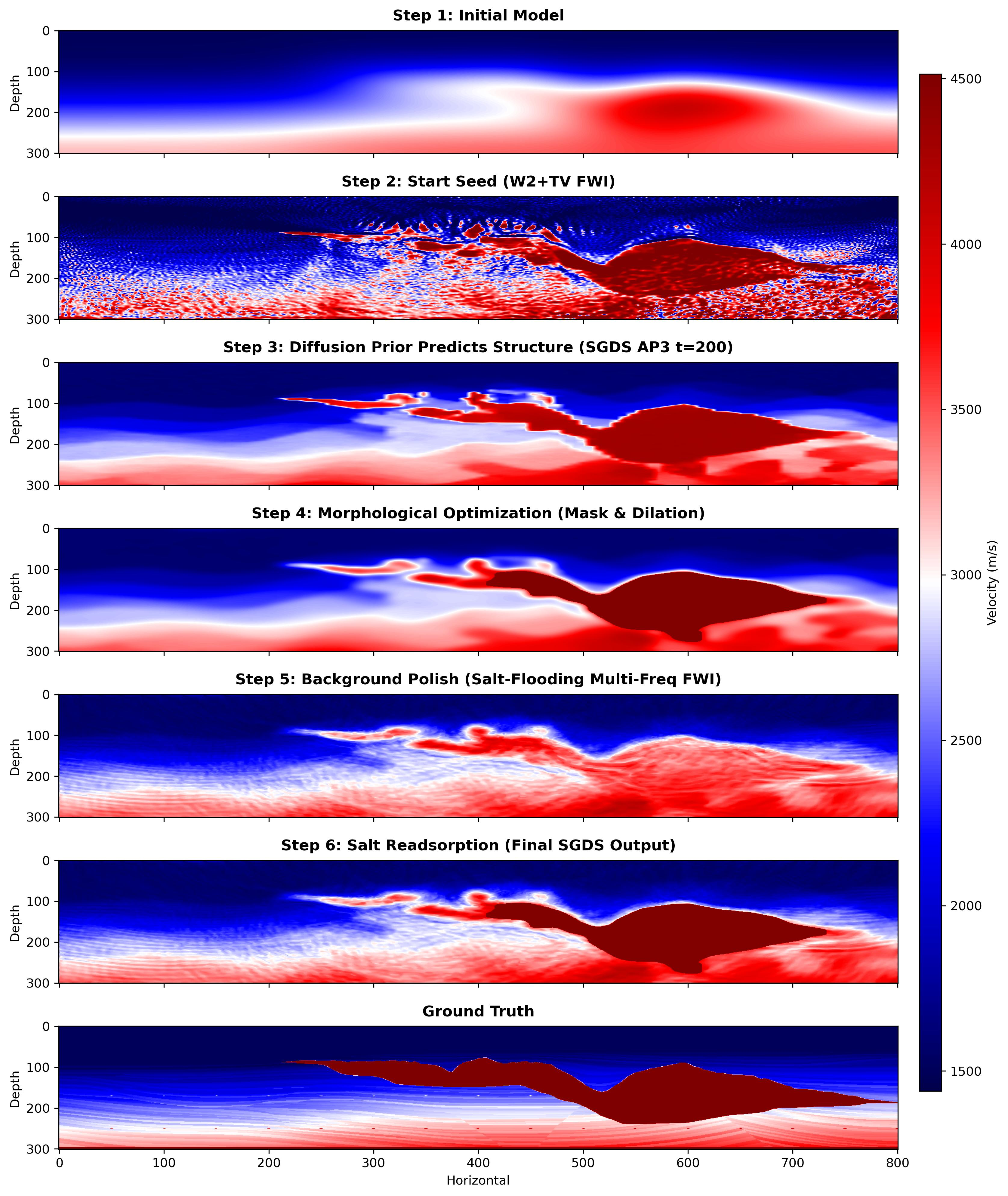}
\caption{Six-stage Sigsbee2A SGDS workflow. From top to bottom, the panels show the smooth initial model, the deterministic seed, the SGDS structural prediction, the morphologically cleaned salt mask, the background-polished model after salt flooding and FWI refinement, and the final reconstruction. This figure clarifies that the difficult salt case uses SGDS as a structural prior within an explicit physics-based workflow.}
\label{fig:sigsbee2a_process}
\end{figure}

\begin{table}[t]
\centering
\caption{Sigsbee2A stress-test metrics. The smooth initial model keeps a high SSIM because it matches the broad background trend, but it does not recover the salt geometry.}
\label{tab:sigsbee_metrics}
\begin{tabular}{lrrrr}
\toprule
\textbf{Method} & \textbf{PSNR} & \textbf{SSIM} & \textbf{RMSE} & \textbf{NRMSE} \\
\midrule
Initial smooth & 13.72 & 0.8060 & 633.5 & 0.2061 \\
FWI $L_2$ & 11.51 & 0.1962 & 817.4 & 0.2659 \\
FWI $L_2$+TV & 13.89 & 0.5732 & 621.6 & 0.2022 \\
W2+TV seed & 13.74 & 0.2431 & 632.0 & 0.2055 \\
Final SGDS-assisted & 16.68 & 0.7182 & 450.8 & 0.1466 \\
\bottomrule
\end{tabular}
\end{table}

These experiments highlight the value of using a learned diffusion prior for synthetic seismic inversion benchmarks. The results support benchmark-scale transfer from GeoFWI patches to Marmousi and Overthrust and motivate field-data and broader out-of-distribution validation as natural next steps. Compared with the classical regularizers tested here, SGDS provides improved resolution and structural accuracy, at the cost of additional diffusion and alternating-update overhead.

\section{Discussion}

Diffusion generative models have recently shown potential for full waveform inversion (FWI), either by conditioning generation on seismic observations or by using pretrained diffusion models as learned priors in reconstruction-based frameworks. However, many existing approaches treat diffusion as an external component rather than integrating it into the classical PDE-constrained optimization loop of FWI. Consequently, the interaction between the physical inversion algorithm and the learned prior can remain indirect, and its effect on stability and convergence behavior requires careful testing.

In parallel, the computer vision community has developed a rich set of diffusion-based methods for inverse problems, including posterior sampling, plug-and-play priors, and guidance-based reconstruction. While these techniques are effective for imaging tasks, fewer studies address strongly nonlinear physical inverse problems governed by wave equations. FWI is a canonical example, characterized by nonconvex optimization landscapes, cycle skipping, and depth-dependent loss of sensitivity. This work aims to bridge these communities by embedding diffusion-based priors directly into the classical FWI framework.

Table~\ref{tab:learning_comparison} summarizes how the present study differs from related learning-based inversion approaches. The table clarifies the scope of the contribution: SGDS uses an unconditional pretrained geological prior and couples it to FWI through alternating likelihood and prior updates, whereas many supervised or conditional methods learn a direct map from paired data or condition the generative model on measurements. A fully controlled ranking against all learned inversion methods would require retraining and retuning each method under the same acquisition, wave solver, and training data.

\begin{table*}[t]
\centering
\caption{Conceptual comparison with representative learning-based inversion strategies.}
\label{tab:learning_comparison}
\small
\setlength{\tabcolsep}{4pt}
\renewcommand{\arraystretch}{1.08}
\begin{tabular}{@{}>{\raggedright\arraybackslash}p{0.22\textwidth}
                  >{\raggedright\arraybackslash}p{0.19\textwidth}
                  >{\raggedright\arraybackslash}p{0.23\textwidth}
                  >{\raggedright\arraybackslash}p{0.28\textwidth}@{}}
\toprule
\textbf{Approach} & \textbf{Training data} & \textbf{Physics coupling} & \textbf{Relation to this work} \\
\midrule
Supervised direct inversion & Paired seismic--velocity data & Learned direct map & Fast but acquisition/distribution tied \\
GAN/VAE priors & Geological model samples & Latent optimization or prior sampling & Learned prior, but less directly tied to FWI likelihood steps \\
Conditional diffusion & Paired or multimodal data & Measurements condition generation & Strong conditioning but requires a data-conditioned generator \\
Diffusion-prior FWI/samplers & Geological prior, sometimes physics guided & Guidance, regularization, or posterior sampling & Closest family; we compare MPGD/SDEdit/SGDS coupling and cost \\
This work & Unconditional GeoFWI prior & Alternating FWI likelihood and denoising & Training-free coupling with benchmark, noise, cost, and stability diagnostics \\
\bottomrule
\end{tabular}
\par\vspace{0.35em}
{\footnotesize Representative examples include supervised direct inversion \cite[]{shucai2020,liubin2021geo}, GAN/VAE priors \cite[]{mosser2020stochastic,laloy2018training}, conditional diffusion \cite[]{wang2024controllable,wang2024seisfusion}, and diffusion-prior FWI/samplers \cite[]{wang2023prior,taufik2024learned,ravasi2025geophysical,chung2022diffusion}.}
\end{table*}

\subsection{Positioning and scope: optimization rather than sampling}

Although diffusion models are naturally connected to Bayesian inference through score-based approximations of probability distributions, our focus is on improving deterministic optimization rather than performing posterior sampling. We study diffusion-guided inversion strategies that incorporate learned score information into iterative FWI updates, encouraging geologically plausible solutions while preserving consistency with seismic observations. In particular, the SGDS workflow alternates between physics-driven gradient updates and prior-driven diffusion steps, providing a practical mechanism for combining physical and learned information.

\subsection{Why diffusion guidance improves FWI}

The numerical results suggest that diffusion guidance is most beneficial in regimes where conventional FWI is poorly conditioned: high-contrast structures (e.g., salt), discontinuities (faults), and deeper regions with limited illumination. In these cases, the seismic gradient may be dominated by reflections and can be insufficient to reliably recover high-wavenumber structure, leading to cycle-skipping artifacts and physically inconsistent updates. Diffusion-based prior steps can complement these limitations by restoring structurally plausible components that are weakly constrained by the data, thereby improving reconstruction fidelity in the tested synthetic settings.

\subsection{Scalability to benchmark-scale models}

A practical challenge in geophysical applications is the mismatch between diffusion training resolutions (typically patch-based) and the scale of realistic velocity models. We therefore treat benchmark deployment as part of the algorithm: direct patching, downsample-single, downsample-multi, and columnwise denoising present different structural scales to the same $100\times100$ prior. The benchmark results show that the deployment choice matters and that simple patchwise use can introduce stitching or scale artifacts. This is why we describe the Marmousi and Overthrust results as benchmark-scale transfer rather than relying on the ambiguous phrase ``zero-shot generalization.''

\subsection{Noise initialization in SGDS}
An ablation study (Figure~\ref{fig:nosie-sgds} and Table~\ref{tab:quantitative}) shows that intermediate diffusion initialization levels provide the best balance between escaping local minima and preserving physically meaningful structure. Excessive noise degrades reconstruction, while insufficient noise limits correction of cycle-skipping artifacts. Developing adaptive schedules based on misfit reduction or uncertainty remains an important direction for future work.

\subsection{Practical deployment considerations}

Several deployment factors determine how the proposed learned-prior workflow should be extended beyond the present synthetic acoustic tests. The training distribution controls the structural vocabulary of the denoiser, and the alternating SGDS loop adds computational overhead relative to conventional FWI. Field-data applications will also require treatment of source-wavelet uncertainty, elastic effects, density variations, anisotropy, and acquisition/modeling mismatch. These considerations point to the next validation stage: multiparameter, elastic, anisotropic, and field-data inversion, together with controlled comparisons against conditional diffusion inversion, DPS/PGDM-style samplers, GAN/VAE priors, and supervised direct inversion networks. Although the fixed-GPU profiling in Table~\ref{tab:cost_accounting} quantifies the current wall-clock and GPU-hour overhead, optimized implementations, batching strategies, and larger benchmark suites may change the absolute timing.

\section{Conclusion}

This work investigates how diffusion generative priors can be integrated into the classical full waveform inversion (FWI) framework. We evaluated three diffusion-guided inversion strategies (MPGD, SDEdit, and Split Gibbs Diffusion Sampling, SGDS) across synthetic geological models and benchmark-scale tests. The results show that learned diffusion priors can improve reconstruction quality compared with conventional $L_2$ and total variation regularization in the tested synthetic settings, particularly in scenarios involving high-contrast structures and limited illumination.

Among the methods considered, SGDS provides the most consistent performance by explicitly alternating between physics-driven likelihood updates and prior-driven diffusion steps. This formulation is consistent with the local alternating-stability diagnostics, mitigates some cycle-skipping artifacts, and improves reconstruction of complex structures, including salt bodies and benchmark models such as Marmousi and Overthrust. These results support diffusion-guided optimization as a practical learned-regularization framework for synthetic and benchmark-scale FWI tests and motivate extension to field-data and broader out-of-distribution settings.

Future work will focus on reducing computational cost, developing adaptive diffusion schedules, and extending the framework to multiparameter, elastic, and field-data inversion settings.

\section*{Data and code availability}
The GeoFWI dataset and associated benchmark code are publicly available through Zenodo at \url{https://zenodo.org/records/17189530} and GitHub at \url{https://github.com/aaspip/geofwi}. Source code for representative SGDS-FWI examples, including reproducibility scripts and a result-preview notebook, is publicly available at \url{https://github.com/shenyiran91/SGDS-FWI}. Large trained checkpoints and intermediate inversion outputs are not included because of file-size limitations, but they can be regenerated using the provided scripts or obtained from the corresponding author upon reasonable request.
\clearpage
\bibliographystyle{plainnat}
\bibliography{example}
\end{document}